\documentclass[12pt,preprint]{aastex}

\usepackage{amsmath}

\begin{document}

\title{Formation of Millisecond Pulsars from Intermediate- and Low-Mass X-ray Binaries}
\author{
Yong Shao and Xiang-Dong Li}

\affil{$^{1}$Department of Astronomy, Nanjing University,
Nanjing 210093, China}

\affil{$^{2}$Key laboratory of Modern Astronomy and Astrophysics
(Nanjing University), Ministry of Education, Nanjing 210093, China}

\affil{$^{}$lixd@nju.edu.cn}

\begin{abstract}
We present a systematic study of the evolution of intermediate-
and low-mass X-ray binaries consisting of an accreting neutron star 
of mass $1.0-1.8 M_{\odot}$ and a donor star of mass 
$1.0-6.0 M_{\odot}$. In our calculations 
we take into account physical processes
such as unstable disk accretion, radio ejection, bump-induced
detachment, and outflow from the $L_{2}$ point. 
Comparing the calculated results with the observations of
binary radio pulsars, we report the following results. 
(1) The allowed parameter space for forming 
binary pulsars in the initial orbital period - donor mass plane  
increases with increasing neutron star mass. 
This may help explain why  some MSPs with 
orbital periods longer than $\sim 60$ days seem to have less 
massive white dwarfs than expected.
Alternatively, some of these wide binary pulsars may be 
formed through mass transfer driven by 
planet/brown dwarf-involved common envelope evolution.
(2) Some of the pulsars in compact binaries might have evolved from 
intermediate-mass X-ray binaries with anomalous magnetic braking. 
(3) The equilibrium spin periods of neutron stars in low-mass
X-ray binaries are in general shorter than the observed spin periods 
of binary pulsars by more than one order of magnitude, suggesting that either the 
simple equilibrium spin model does not apply, or there are other 
mechanisms/processes spinning down the neutron stars.

\end{abstract}

\keywords{stars: millisecond pulsars $-$ stars: evolution $-$ MSPs: general}

\section{Introduction}

Millisecond pulsars (MSPs) are neutron stars (NSs)
characterized by short
spin periods ($P_{\rm spin} < 20 $ ms) and weak surface
magnetic fields ($B < 10^{9}$ G), which are often found in
binaries with a white dwarf (WD) companion. It is generally
agreed that MSPs are old NSs,
recycled by accretion of mass and angular momentum from
the donor stars through Roche-lobe overflow (RLOF) during the previous
low-mass X-ray binary (LMXB) evolution. The NS was spun
up, due to mass accretion, into a MSP, while the donor evolved
to be a He or CO WD \citep[see e.g., ][for reviews]{bv91,tv06}.

%In the past, for the study of binary population synthesis,
%the criteria for dynamically unstable mass transfer was
%considered during the evolution of X-ray binaries, and some
%analytic approximations were adopted\citep[e.g.,][]
%{rjw82,lvv95,kw96,kr99}.

The stability of mass transfer in an X-ray binary
depends on the ratio of the masses of the donor star and
the NS, and the initial orbital period of the system.
Traditionally, it was thought that, if the
mass ratio is large enough ($\gtrsim 1.5$), the mass
transfer is likely to be unstable, resulting in a common envelope (CE)
evolution \citep{p76,w84,il93}, where the NS
spirals into the envelope of the donor on a
very short timescale ($<10^{3}$ yr). 
More recent investigations \citep[e.g.,][]{tvs00,p00,kdkr00,prp02,prp03}
show that  X-ray binary systems with
intermediate-mass (up to $\sim 5 M_{\odot}$) donor stars
(i.e., IMXBs) can avoid the spiral-in phase and experience
rapid mass transfer on a thermal timescale, 
successfully evolve to become
LMXBs.

For LMXBs, a CE evolution is even unavoidable if the separation of
binary components is large enough, so that the donor reaches the asymptotic
giant branch (AGB) phase with a deep convective envelope 
before RLOF (i.e. Case C RLOF).
Stable mass transfer in LMXBs usually occurs on a timescale
$\sim 10^{8}-10^{10}$ yr when the donor star is
on the main sequence or (sub)giant branch at the onset
of RLOF (i.e., Case A or B RLOF),
and forming MSPs seems to be feasible
in this process. 
It was found by \citet{ps88,ps89} that there exists a critical
bifurcation orbital period ($P_{\rm bif}$), 
the initial orbital period that
separates the formation
of converging LMXBs (which evolve with
decreasing orbital periods until the donor star
 becomes degenerate or an ultra-compact
binary is formed) from diverging LMXBs. The
value of $P_{\rm bif}$ is found to be $\sim 1$
d, but depends heavily on the processes of tidal
interactions and the mechanisms and efficiency of 
orbital angular momentum loss \citep{e98,prp02,vs05,ml09}.

%y it was indicated that many,
%perhaps most current LMXBs may have evolved from
%intermediate-mass ($\gtrsim$ 2.0 $M_{\odot}$)
%X-ray binaries (IMXBs) .
%These authors have performed systematic studies of
%the evolution of L/IMXBs, which include a donor of mass from
%$\lesssim1\,M_{\odot}$ to $\sim 6.0\,M_{\odot}$
%and a canonical NS. The initial
%orbital periods $P_{\rm orb}$ covered the interval from less than 1 d to
%several hundred days.
%The binary evolution calculations showed that many

The final products of LMXB evolution are binary pulsars.
The distributions of the spin periods of the pulsars, the orbital
periods and the WD masses can be used to testify
the models of I/LMXB evolution.
\citet{d08} compared the theoretical expectations of
I/LMXB evolution to the populations of Galactic binary pulsars.  
He showed that a significant population
of binary pulsars with 1 d $\lesssim P_{\rm orb} \lesssim 100$ d
are generally consistent with being the
descendants of long-period LMXBs or IMXBs.
However, there remain quite a few unresolved puzzles.
For example, binary pulsars with $P_{\rm orb}\gtrsim$ 60 d seem 
to have WD companions less massive
than predicted by theory,
as pointed out previously by \citet{ts99}, and those
with 0.1 d $\lesssim P_{\rm orb}\lesssim$ 1 d are inconsistent
with any I/LMXB evolution.

In the previous studies on I/LMXB evolution, a canonical NS
(of mass $\sim 1.3-1.4\,M_{\odot}$)
was usually adopted. This seems to be supported
by the finding that the NS mass distribution 
is consistent with a narrow Gaussian at
$1.35\pm 0.04\,M_{\odot}$ \citep{tc99}.
However, both
observations \citep[see][and references therein]{z11,k10,spr10,o12} and
theories \citep{n84,tww96,h03,wj05} suggest that the initial masses
of NSs may occupy a large range, probably
originating from two different mechanisms of
forming NS: iron-core collapse supernovae and
electron-capture supernovae.
NSs in  high-mass X-ray binaries (HMXBs)  have experienced
very little accretion because of their young ages, so their masses should be
very close to those at birth.
The measured masses of NSs in HMXBs range from $1.06^{+0.11}_{-0.10}
\,M_{\odot}$ for SMC X$-$1 \citep{v07} to
$1.86\pm 0.16\,M_{\odot}$ for Vela X$-$1 \citep{b01,q03}.
Recently \citet{r11} present an improved method for determining the mass of
NSs in eclipsing X-ray pulsar binaries and apply it
to six systems. They find that the NS masses  range from 
$0.87\pm 0.07\,M_{\odot}$
(eccentric orbit) or $1.00\pm0.10\,M_{\odot}$ (circular orbit)
for 4U1538$-$52 to $1.77\pm 0.08\,M_{\odot}$ for Vela X$-$1.
So in a proper investigation, the influence of the NS masses 
should be included. Recent evolutionary calculations by 
\citet{db10} have provided evidence
that the evolution of I/LMXBs
depends upon the NS mass. In this work we perform
systematic calculations of I/LMXB evolution and discuss
the properties of the produced binary and millisecond pulsars (BMSPs), taking into 
account different initial NS masses. We adopt the initial donor
masses to be $1.0 - 6.0 M_{\odot}$, and two
different initial masses (1.0 and 1.8 $M_{\odot}$) for the NS.

During the mass transfer processes, part of the
transferred mass from the donor star may escape
from the binary system, carrying away the orbital
angular momentum. The formation of BMSPs is closely
related to the mechanisms of mass and angular momentum
loss, and the stability of mass transfer is also
dependent on the angular mementum loss rate
\citep{spv97}.
Generally, it is assumed that all the mass transferred
from the donor will accrete onto the NS unless in a 
super-Eddington mass transfer phase, during which the NS
will accrete at the Eddington rate 
($\dot{M}_{\rm Edd}\sim 1.5\times 10^{-8}\,M_{\odot}$\,yr$^{-1}$
for a $1.4\,M_{\odot}$ NS), and the residual mass
escapes from the binary system, carrying the
NS's specific orbital angular momentum. This is
the so-called ``isotropic reemission model''.
Actually, there may also be mass loss even during the
sub-Eddington mass transfer phase. For example,
the accretion disk can become thermally
and viscously unstable
when the orbital period is larger than a critical value
\citep{v96,dlhc99}, leading to limit cycle behavior of the mass
transfer rate.
The NS can accrete mass only 
during outbursts, while most matter may be ejected
out of the binary systems during quiescence by the 
radiation and magnetic pressure of the rapidly rotating
NS  \citep{rst89,bdb02}.
In our calculations, mass loss due to super-Eddington
mass transfer, radio ejection caused by an unstable disk
\citep{bdb02} or bump-induced detachment
\citep{dvl06}, and in some cases outflow from the $L_{2}$
Lagrangian point  are included.

This paper is organized as follows. In Section 2
 we describe the stellar evolution code and the
binary model used in this paper. We present the calculated results
in Section 3 and discuss their possible applications
in the formation of BMSPs in Section 4,
and summarize in Section 5.

\section{Binary Evolutionary Calculations}

\subsection{The Stellar Evolution Code}

All calculations were carried out with an updated
version of the stellar evolution code developed
by \citet{e71,e72} \citep[see also][]{h94,p95}.
We set initial solar chemical compositions
(i.e., $X = 0.7$, $Y = 0.28$, and $Z = 0.02$) for the
donor star, and take the ratio of the mixing length to
the pressure scale height to be 2.0, and the general
convective overshooting parameter to be 0.12. In our
calculations, we have considered a number of
binary interactions in order to 
follow the details of the mass transfer process,
including orbital angular momentum
loss due to gravitational radiation (GR),
magnetic braking (MB), mass loss, and the effect of disk
instability.

\subsection{ The Input Physics}

Each of the binaries initially consists of an NS
of mass $M_{1}$ ($=1.0$ or
1.8 $M_{\odot}$) and a zero-age main-sequence (ZAMS) donor of
mass $M_{2}$ ($\sim 1.0-6.0 M_{\odot}$). The efective
radius of the Roche lobe $R_{L}$ of the donor is given by the
following formula  \citep{e83},
\begin{equation}
R_{L}/a = \frac{0.49q^{-2/3}}{0.6q^{-2/3} + \ln(1 + q^{-1/3})},
\end{equation}
where $a$ is the orbital separation of the binary,
and $q = M_{2}/M_{1}$ is the mass ratio of the
binary components.
As usual we assume that tides keep the binary orbit
circular \citep{k88}, so that the orbital angular momentum is
\begin{equation}
J_{\rm orb} = \frac{M_{1}M_{2}}{M}\Omega a^{2},
\end{equation}
where $M =M_{1}+M_{2}$, and $\Omega=\sqrt{GM/a^{3}}$ is
the orbital angular velocity. Logarithmic
differentiation of Eq.~(2) with time gives the rate of
change in the orbital separation
\begin{equation}
\frac{\dot{a}}{a} = 2\frac{\dot{J}_{\rm orb}}{J_{\rm orb}}-
2\frac{\dot{M_{1}}}{M_{1}}-2\frac{\dot{M_{2}}}{M_{2}}+\frac{\dot{M}}{M},
\end{equation}
where the total changing rate in the orbital angular momentum is
\begin{equation}
\frac{\dot{J}_{\rm orb}}{J_{\rm orb}} =\frac{\dot{J}_{\rm GR}}{J_{\rm orb}}+
\frac{\dot{J}_{\rm MB}}{J_{\rm orb}}+
\frac{\dot{J}_{\rm ml}}{J_{\rm orb}}.
\end{equation}
The three terms on the right-hand-side of Eq.~(4) represent
angular momentum loss due to GR, MB,
and mass loss, respectively.
The rate of angular momentum loss due to GR
is calculated according to the standard
formula \citep{ll59,f71}
\begin{equation}
\frac{\dot{J}_{\rm GR}}{J_{\rm orb}}=
-\frac{32G^{3}}{5c^{5}}\frac{M_{1}M_{2}M}{a^{4}},
\end{equation}
where $G$ and $c$ are the gravitational constant and
the speed of light, respectively.
The prescription of \citet{vz81} is adopted to 
calculate the angular momentum loss due to
MB, 
\begin{equation}
\frac{\dot{J}_{\rm MB}}{J_{\rm orb}}=
-3.8\times10^{-30}\frac{GR_2^{4}M^{2}}{a^{5}M_{1}} \,{\rm s}^{-1},
\end{equation}
where $R_2$ is the radius of the donor.
%and the angular
%velocity $\omega$ of the secondary is assumed
%to be synchronized with the orbit ($\omega=\Omega$).

For IMXBs and wide LMXBs, the mass transfer rate 
$|\dot{M_{2}}|$ (note that $\dot{M}_2<0$) can be larger than the 
Eddington accretion rate 
$\dot{M}_{\rm Edd}$. Thus the
mass loss rate from the binary system is $\dot{M} =\dot{M_{2}}+
\dot{M}_{\rm Edd}$. Here, we adopt the isotropic reemission
model, assuming that the extra material leaves the binary in the form of
isotropic  wind from the NS, carrying  off
the NS's specific  orbital angular momentum 
$j_{1}=(M_{2}/M_{1}M)J_{\rm orb}$.  The related angular momentum loss 
rate can be derived to be
\begin{equation}
\frac{\dot{J}_{\rm ml,1}}{J_{\rm orb}}=
\dfrac{M_{2}}{M_{1}M}(\dot{M}_{2}+\dot{M}_{\rm Edd}).
\end{equation}

In some cases part of the material lost from the donor star
may escape the system through the $L_{2}$ Lagrangian point 
rather accrete onto the NS.
Assuming that a fraction $\delta$ of the matter flow escapes
from the binary with the specific orbital angular momentum
$j_{2}=a_{L_{2}}\Omega$,
the angular momentum loss rate due to the $L_{2}$
point outflow is given by \citep{sl12}
\begin{equation}
\frac{\dot{J}_{\rm ml,2}}{J_{\rm orb}}=\delta\frac{a_{L_{2}}
^{2}}{a^{2}}\frac{M\dot{M}_{2}}{M_1M_2},
\end{equation}
where $a_{L_{2}}$ is the distance between the mass
center of binary and the $L_{2}$ point.

The transferred material from the donor will form
an accretion disk surrounding the NS. If the effective temperature
in the accretion disk is below $\sim6500$ K (the
hydrogen ionization temperature), the  accretion disk is likely to be
thermally and viscously unstable \citep{l01}. In LMXBs 
irradiation from the
NS may help stabilize the disk to some extent
\citep{v96,k97,l01,r08}. The critical mass transfer
rate for the disk instability is given by \citep{dlhc99}
\begin{equation}
\dot{M}_{\rm cr}\simeq 3.2\times10^{-9}(\frac{M_{1}}{1.4M_{\odot}})^{0.5}(\frac{M_{2}}
{1.0M_{\odot}})^{-0.2}(\frac{P_{\rm orb}}{1.0\,{\rm d}})^{1.4} M_{\odot}\, {\rm yr}^{-1}.
\end{equation}
If the mass transfer rate is less than $\dot{M}_{\rm cr}$,
the X-ray binary is assumed to become transient,
experiencing short outbursts separated by 
long quiescent intervals.
Material accumulates in the disk during the quiescent 
phase while the NS
accretes mostly during outbursts. Define the duty
cycle $d$ \citep[$\sim 0.001-0.1$, see][]{k03}
to be the ratio of the outburst timescale to the recurrence
time, the accretion rate during outbursts can be estimated
as $\sim - \dot{M}_{2} /d$ . Of course, the mass
accretion rate in this case is also limited by the Eddington
accretion rate.

\section{ Results of I/LMXB Evolution Calculations}

%Here, we obtain a series of correlations from the binary
%evolution, which is vary similar with the work of \citet{ts99,tvs00,prp02}
%in expression forms, such as the $P_{orb}^{final}-M_{WD}$
%relation, and the Initial ( $M_{2} , P_{orb}$ ) Parameter Space
%to form MSPs with WD components. In our study, we adopt
%two different masses of neutron star: 1.0 and 1.8 $M_{\odot}$, the
%results will be presented respectively.

\subsection{The Initial $M_{2}-P_{\rm orb}$ Parameter Space for
Successful Evolution}

In Fig.~1 we outline the results of our calculations
by showing the fate of the I/LMXB evolution in the initial
$M_{2}-P_{\rm orb}$ diagram, similar as in \citet{tvs00}.
The left and right panels correspond to $1.0\,M_{\odot}$ and
$1.8\,M_{\odot}$ NS, respectively. It is seen that the allowed
space for successful evolution into binary pulsars (i.e.,
without CE evolution) is larger
for $M_1=1.8\,M_{\odot}$ than for $M_1=1.0\,M_{\odot}$,
since the lower mass ratio in the former case can 
stabilize mass transfer during the IMXB evolution.
When $M_1=1.0\,M_{\odot}$,
systems with a low mass companion ($M_{2}\lesssim 1.3 M_{\odot}$)
can form binary radio pulsars with a He/CO WD companion
if the initial orbital period is $<1000$ d,
while the lower limit of $M_{2}$ increases to
$\sim 2.2\,M_{\odot}$ when $M_1=1.8\,M_{\odot}$.
IMXBs can avoid spiral-in and CE evolution
when $1.5M_{\odot}\lesssim M_{2}\lesssim 3.3M_{\odot}$
and $2\, {\rm d}\lesssim  P_{\rm orb} \lesssim 10$ d for $M_1=1.0\,M_{\odot}$,
and $2.4M_{\odot} \lesssim M_{2}\lesssim 5.5M_{\odot}$ and
$2\,{\rm d}\lesssim P_{\rm orb} \lesssim 40$ d
for $M_1=1.8\,M_{\odot}$, respectively. 
In both cases, if the initial $P_{\rm orb}$
is too short, the binary
systems will experience either a CE phase, or become X-ray binaries
with degenerate hydrogen stars \citep{e98}, possibly
forming ultra-compact binaries. On the other hand, if the initial
$P_{\rm orb}$ is too long, the donor will develop
a deep convective envelope at the onset of RLOF, and a runaway mass transfer
is initiated, leading to CE evolution. In \citet{tvs00}
systems with $M_1=1.3\,M_{\odot}$,
$2M_{\odot} \lesssim M_{2}\lesssim 5M_{\odot}$
and $1\,{\rm d}\lesssim P_{\rm orb} \lesssim 20$ d can survive
unstable mass transfer. These systems are right
 between the limiting cases
presented in Fig.~1.

%\subsection{a case of $M_{2}$ = 4 $M_{\odot}$, $P_{orb}$ = 10 days }
To see in more detail how the NS mass affects the binary evolution,
we show in Fig.~2 the evolutionary paths of two IMXBs
with the same donor mass $M_{2} = 4\,M_{\odot}$ and orbital period
$P_{\rm orb}$ = 10 d, but different NS mass
(1.0 $M_{\odot}$ and 1.8 $M_{\odot}$).
At the age of $\sim 175.1$ Myr, the donor star starts to 
transfer mass via RLOF. In the upper two panels with a 1.0
$M_{\odot}$ NS, mass transfer occurs rapidly, rising up to
$>10^{-4}M_{\odot}$\,yr$^{-1}$ within $\sim 10^4$ yr.
Meanwhile, the orbital period reduces to $\lesssim 4$ d.
This dynamically unstable mass transfer will result in 
CE evolution and probably merging of the NS and the
He core of the donor.
In the lower two
panels with a 1.8 $M_{\odot}$ NS, mass transfer initially
proceeds at a rate of $\sim3\times10^{-5}M_{\odot}$\,yr$^{-1}$
for $\sim 10^{5}$ yr, during which most of the hydrogen-rich 
envelope ($\sim 3
\,M_{\odot}$) is removed from the donor star. 
The mass ratio inverts during this phase,
and subsequently the mass transfer rate decreases
to be several $10^{-7}M_{\odot}$\,yr$^{-1}$ for
$\sim 10^{6}$ yr.  During this time the NS  accretes about $0.012
M_{\odot}$ mass, and this is the primary process to spin up the
NS. The final product is a binary consisting of a $1.815
\,M_{\odot}$ (mildly) recycled NS and a $\sim 0.6M_{\odot}$
CO WD in an orbital period of $\sim36$ d.

\subsection{The $P_{\rm orb}^{final}-M_{\rm WD}$ Diagram}

Figure 3 shows the calculated correlation between the final orbital
period $P_{\rm orb}^{final}$ and the mass of the WD (the remnant of the 
donor) $M_{\rm WD}$, for two different initial NS masses, $1.0 
M_{\odot}$ (left panel) and $1.8 M_{\odot}$ (right panel). 
As mentioned above, low-mass
donor stars ($M_{2}\la 1.3M_{\odot}$ or $\la 2.2 M_{\odot}$ for
$1.0 M_{\odot}$ or $1.8 M_{\odot}$ NS, respectively)
may evolve to be He WDs.
However, when the mass of He is accumulated to exceed
$\sim 0.4-0.5M_{\odot}$ in the core of the donor, He flash
will occur, giving rise to the formation of a CO core. The orbital period
will reach $\ga 10^{3}$ days. 
Intermediate-mass donors can avoid
He flash due to their higher temperature, forming CO WDs 
with the mass $M_{\rm WD}\gtrsim 0.33\,M_{\odot}$.
Their $P_{\rm orb}^{final}-M_{\rm WD}$ distribution 
deviates from that of the low-mass branch obviously. 
%Only
%systems with proper $P_{\rm orb}$ (see Fig.~1)
%can avoid CE evolution, if the donor stars
%have a radiative envelope or a thin convective envelope.

\subsection{Radio Ejection during the ``Bump-related" Detachment?}

When its hydrogen shell reaches a discontinuity
in the hydrogen content at the time of the first
dredge-up, a low-mass star will suffer
a temporary contraction, thus producing a
``bump'' in the luminosity function of the red giants.
\citet{dvl06} suggested that the orbital period gap
($\sim 20 -60$ days) of BMSPs may be 
%due to the massloss during 
related to the bump-induced
detachment of the donor star from its RL. 
If the NS has already accreted sufficient mass,
it may turn on as a MSP during the detachment phase.
Material transferred from the
donor, once it expands
again to refill its Roche lobe, may be inhibited by the pulsar's
radiation pressure and ejected at the inner lagrangian point $L_{1}$
\citep{rst89,bdb02},  so that no further mass accretion
would occur. Similarly, this so-called ``radio ejection" process
may also occur in LMXBs with an unstable accretion
disk \citep{bdb02}. In both cases the mass loss will influence 
the orbital evolution of the binary and the spin evolution
of the NS. \citet{dvl06} considered the radio ejection only 
during the bump-related detachment. Our calculations show that the disk
instability generally appears earlier than the bump phase. 
Two examples are presented in Fig.~4. In the upper two panels, the 
binary system initially consists of
a 1.0 $M_{\odot}$ NS and a 1.0 $M_{\odot}$
ZAMS donor star in an orbital period of 5.0 d. RLOF 
initiates  at the age $\sim 1.186\times10^{10}$ yr, and the bump
occurs $\sim7\times10^{7}$ yr later, at which the donor mass has
decreased to be $\sim 0.52
M_{\odot}$. However, the disk becomes thermally unstable when the
donor mass is $\sim 0.9M_{\odot}$, much earlier than the
bump phase. If the NS's spin was accelerated during the prior mass
transfer, the accumulated mass in the disk during
quiescence will be ejected out of
the binary by the radiation pressure of the rapidly spinning
NS. We find a similar situation in the case of 
a binary containing a 1.8
$M_{\odot}$ NS and a 1.5 $M_{\odot}$ donor star shown
in the lower two panels.
Compared with the effect of unstable disk accretion,
it seems that the bump-related detachment 
might play a less important role in the evolution of LMXBs.

We finally note that the process of pulsar-driven mass ejection is
highly uncertain, since its condition and efficiency depend on several
unknown parameters \citep{fl11}, such as the value of the equilibrium period
that a NS will reach during the mass transfer. However, as
seen below, there is controversy on the estimate of the
equilibrium period for NSs in LMXBs, and one should be 
cautious when considering the effect of radio ejection
on the binary evolution.

\subsection{The effect of outflow from the $L_{2}$ point}

In the binary evolution the stability of mass transfer 
strongly depends on mass loss and related angular 
momentum loss. Although we have taken into account
 various ways for mass loss, including super-Eddington 
 mass transfer, unstable mass
transfer due to disk instability, and mass ejection due to
the propeller effect and the pulsar's radiation,
the detail processes are  complicated
and uncertain, and  simplified treatment
might not reveal the realistic situation.
In the following, we show an example of how outflows from the 
binary systems through the $L_{2}$ point influence the evolution of
mass transfer, and change the conditions of forming BMSPs.

MSP J1614$-$2230 is a 3.15 ms pulsar of mass 
1.97 $\pm$ 0.04 $M_{\odot}$ \citep{d10}. The mass of 
its WD companion ($0.500 \pm 0.006M_{\odot}$) and the 
orbital period (8.7 d) were also measured accurately.
\citet{l11} and \citet{tlk11} systematically investigated 
the formation channels of this pulsar from an IMXB. They showed 
that NS this massive are not easy to produce 
in spite of the initially high mass of the donor star, 
unless they were already born as a relatively massive NS.
However, \citet{tlk11} found that, for the system with a 1.8 $M_{\odot}$ NS,
the final orbital period will be always larger than 10
d, inconsistent with the observation of the pulsar.
This conclusion will not hold if we let a small fraction $\delta$
of the transferred mass leave the system from the $L_{2}$ point
\citep[e.g.,][]{b89}.
As an illustration,
in Fig.~5 we compare the evolution of an IMXB containing  
a 1.8 $M_{\odot}$ NS and a 4.5 $M_{\odot}$ companion 
star in an orbit of 2.4 d, with $\delta = 0$ (upper two panels)
and $\delta = 0.04$ (lower two panels), respectively.
In the former case the primary angular momentum
loss mechanism is the isotropic reemission around the NS
during the super-Eddington accretion phase. 
The final  system consists of a 2 $M_{\odot}$ NS and 0.5 $M_{\odot}$
CO WD, with a orbital period 15 d. In the latter case 
angular momentum loss due 
to the $L_{2}$ point outflow plays a role when the system
evolves into an LMXB, and reduces the orbital period to $\sim 8.7$ days.

\section{Comparison with BMSPs}

\subsection{The $P_{\rm orb}-M_{\rm WD}$ relation for low-mass BMSPs}

In wide LMXBs, the donor will climb to
the red-giant branch (RGB) in the HR diagram before RLOF.
For low-mass stars ($< 2.3 M_{\odot}$) on the RGB,
there is a well known relationship between the mass of the degenerate
He core and the radius of the giant star, which is almost
entirely independent of the mass of the hydrogen-rich envelope
\citep{rw71,w83}. Based on this relationship, a specific
correlation between the orbital period $P_{\rm orb}$ 
and the WD mass $M_{\rm WD}$ is obtained
\citep{j87,r95,ts99}. Comparison with the observations
shows that a significant population of BMSPs with He WD
companion is generally 
consistent with this $P_{\rm orb}-M_{\rm WD}$ relation.
However, there seems to be a systematic deviation 
from the correlation for
pulsars with $P_{\rm orb} \gtrsim 60$ d, which 
seem to have WD companions lighter than expected
\citep{t96}. 

Both systematic small values of the orbital inclination
$i$ and large NS mass can increase $M_{\rm WD}$
for the given observed mass functions.
Since there does not seem to be any observational selection 
effect favoring small inclination angle $i$ \citep{t96}, we first examine
whether the $P_{\rm orb}-M_{\rm WD}$ correlation can be accounted
for if the long-period BMSPs have CO WD 
companions. For example,
\citet{s05} noticed that the \citet{ts99} $P_{\rm orb}-M_{\rm WD}$ 
relation is incompatible at the $99.5\%$ level  with
a uniform distribution of $\cos i$ if the pulsar masses are drawn
from a Gaussian distribution centered on $1.35 M_{\odot}$
with width $0.04 M_{\odot}$, and better agreement with 
uniformity in $\cos i$ can be reached if the pulsar masses are 
large on average (e.g., $1.75\pm 0.04M_{\odot}$).
An extreme example is PSR B0820$+$02, which has 
a $0.6\,M_{\odot}$ CO WD 
companion in a very wide orbit with $P_{\rm orb}\simeq 1232$ d
\citep{kr00}. However, in such  wide binaries
it is difficult for the NS to accrete  enough matter
\citep{lw98,ts99}, and the NS must be born
heavy if this interpretation is correct. In Fig.~6 we compare the 
relations between 
$P_{\rm orb}$ and $M_{\rm WD}$
in the cases that the NS has an initial mass of 1.0 $M_{\odot}$
(left panel) and 1.8 $M_{\odot}$ (right panel).
Also plotted are binary pulsars with measured $P_{\rm orb}$ 
and $M_{\rm WD}$ (90\% probability
mass range for randomly oriented orbits) for a fixed NS mass 
of 1.2 and
2.0 $M_{\odot}$, respectively (i.e., in each case the NS is 
assumed to have accreted  $0.2 M_{\odot}$
during the mass transfer). 
It is seen that some binary pulsars with $P_{\rm orb} >$ 100 d can fairly
match the relation if they have massive NSs 
($\sim2M_{\odot}$) and heavy
CO WDs (although in some cases small orbital inclination angles
may be required),
while for those with $P_{\rm orb} <$ 20 d, statistically lighter NSs 
($\sim1.2M_{\odot}$)
seem to follow the relation better. 

It is expected that the NS has to accrete at least  a few $\sim 
0.01\,M_{\odot}$ mass to reach a millisecond period, but
this is difficult to achieve for NSs in wide binaries \citep[see also][]{lc11}, 
because the mass transfer rate (which increases with increasing
orbital period) is likely to be super-Eddington,
and the accretion disk is likely to be unstable.
In Fig.~7 we show the mass transfer rate
$|\dot{M}_{2}|$, and the accreted mass $\Delta M_{1}$
of the NS as a function of the final
orbital period $P_{\rm orb}$. 
Note that here we plot the mass
transfer rate only for the stable mass transfer phase -
when the accretion disk becomes unstable,
we take it to be the critical value $\dot{M}_{\rm cr}$
at the onset of disk instability.
Hence both the mass accretion 
rate and $\Delta M_{1}$ are limited by the value 
of $\dot{M}_{\rm cr}$: when 
$|\dot{M}_{2}|$ is less than the $\dot{M}_{\rm cr}$,
unstable disk accretion occurs, and the
NS is assumed to accrete mass only during outbursts, 
and part of the matter will be ejected out of the binary
systems if $|\dot{M}_{2}/d|$ is super-Eddington.

The left panel of  Fig.~7 shows the results for a 
$1.0\,M_{\odot}$ NS
with a companion star of initial mass $M_2=1.0$, 2.0 and 3.0
$M_{\odot}$, respectively. Generally more massive
donor stars result in higher mass transfer rate,
which also increases with longer orbital
period $P_{\rm orb}$. 
It is noted that (1) $|\dot{M}_{2}|$
is always $\gtrsim 0.1\dot{M}_{\rm Edd}$,
and (2) $\Delta M_{1}\lesssim 0.3 M_{\odot}$, 
which decreases with $P_{\rm orb}$, because in wider 
systems the mass accretion rates (with both  stable and 
unstable accretion disks) are more likely to be super-Eddington.
The right panel is for the systems with a $1.8 M_{\odot}$ 
NS. In the case of  light companion star
($\sim 1.0 M_{\odot}$), the initial orbital period should be larger 
than $\sim1$ d, so that the critical mass transfer
rate $\dot{M}_{\rm cr}$ is always larger than the
mass transfer rate, since $\dot{M}_{\rm cr}$
increases during the mass transfer process
(with increasing orbit period),
while the mass transfer rate $|\dot{M}_{2}|$ decreases
all the way \citep{w83}.
The accreted mass $\Delta M_{1}$ by the NS is thus
significantly lower than
those with more massive donors.
So only results with  2.0 and 3.0
$M_{\odot}$ donor stars are presented, 
in which $\Delta M_{1}\lesssim
0.35 M_{\odot}$, and $\dot{M}_{2}\gtrsim
0.3 \dot{M}_{\rm Edd}$.

Figure~7 shows that in general 
$\dot{M}_{2}\gtrsim 0.1 \dot{M}_{\rm Edd}$,
and $\Delta M_{1} \lesssim 0.3 M_{\odot}$.
The  distribution of the mass transfer rate is considerably 
higher than that ($\dot{M}_{2} \sim10^{-11}-10^{-8} M_{\odot}$\, yr$^{-1}$)
obtained by \citet{prp02}, but more compatible with 
the observations of persistent LMXBs. The main reason is that
we have taken into account
the effect of disk instability.  The $\Delta M_{1}$ distribution
is roughly in line with \citet{lc11},
who found that $\Delta M_{1}$ is generally
less than 0.6 $M_{\odot}$ in their calculations for
systems  with a
1.4 $M_{\odot}$ NS and a $1.0-2.0 M_{\odot}$
donor star. Moreover, we find that systems with initially massive 
NSs ($1.8\,M_{\odot}$) 
may accrete enough mass to evolve into BMSPs
 with 60 d $\la P_{\rm orb}\la 200 $ d, , while light
NSs in wide binaries are more likely to be partially 
recycled. This seems to
support our conjecture that some of the wide BMSPs 
might be born massive. Hopefully accurate measurements of both the 
pulsar and the WD masses will help settle this issue.

\citet{l05} performed timing observations of the 
BMSP J1640$+$2224 (with an orbital period of 
175 d), and constrained the WD mass to be
$0.15^{+0.08}_{-0.05}\,M_{\odot}$
($1 \sigma$ uncertainties), which indicates that
the companion is very likely to be a low-mass He WD.
In this case the massive NS + CO WD model obviously
does not work, and one has to explore other possibilities.
Evaporation of the companion star from a
wind of relativistic particles after the pulsar turns on may
decrease the companion mass significantly, but it is unlikely
for wide BMSPs, since the evaporation timescale
would be longer than the Hubble time
\citep{t96}. 

Here we suggest another possible solution.
It is interesting to note that solar-type stars  are 
usually found to be surrounded by sub-stellar companions 
(usually planets and/or brown dwarfs) \citep{c12}.
One may expect  that in some relatively wide 
LMXBs the companion star had
possessed substellar companion(s) in close orbits
like ``hot Jupiters".
When the star evolved on the giant branch it would become
big enough to capture its planet/brown dwarf.
The planet/brown dwarf spiraled into the envelope of the 
giant to initiate a CE phase. The
frictional drag arising from its motion through the
CE would lead to loss of its orbital angular 
momentum and deposit of orbital energy in the envelope.
If there was enough orbital energy, the spiral-in process would
expel the envelope of the giant, leaving a WD remnant 
\citep{nt98}\footnote{In addition, \citet{bs10} suggest that the
binary systems may reach 
stable synchronized orbits before the 
onset of the CE phase. Such stable synchronized orbits allow the 
RGB star to lose mass prior to the onset of 
the CE phase. Even after the secondary enters the giant envelope, 
the rotational velocity is high enough to
cause an enhanced mass-loss rate.}.
If the initial separation between the star and the substellar
object(s) is less than tens of Solar radii, 
the final outcome would be an under-massive WD 
with or without the surrounding planet/brown dwarf, depending on
whether it evaporated, filled its own Roche-lobe, or survived. 
The discovery of a $0.053 M_{\odot}$ brown dwarf 
in a short (0.08 d) period orbit 
around a $0.39 M_{\odot}$ WD 0137$-$349 \citep{m06} presents
strong observational evidence for this interaction.

During the planet/brown dwarf-involved CE phase, the companion's 
envelope expanded rapidly and filled its Roche-lobe,
leading to mass transfer onto the NS. Since this phase
was very short ($<10^3$ yr), the mass transfer rate would be 
much higher than the Eddington limit rate for the NS,
so that the NS accreted very small mass $\sim 10^{-5} M_{\odot}$ 
(unless the CE phase lasted much longer time). 
If this is the case, we have to require that
{\em some MSPs were born this way, rather recycled during the
LMXB evolution}. In the literature, this idea has already been discussed
by \citet{mh01}, who showed that, the existence 
of the innermost, moon-sized planet in the PSR 1257$+$12 system 
suggests that the pulsar was born with 
approximately its current spin frequency and magnetic field.
A schematic view of the formation of BMSPs
with planet/brown dwarf-involved CE evolution is shown in Fig. 8.

Not only having an impact on the formation of MSPs, 
the substellar objects,
if really exist in low-mass binaries, may also
play a role in the CE evolution, especially influence the
estimate of the CE efficiency parameter $\alpha_{\rm CE}$. 
Recently \citet{dkk12} reconstructed  
the CE phase for the current sample 
of post-CE binaries (PCEBs) with observationally 
determined component masses and orbital periods.
Searching for correlations between $\alpha_{\rm CE}$ and the 
binary parameters, they found that,  when the 
internal energy of the progenitor primary envelope is taken into 
account,  $\alpha_{\rm CE}$ 
decreases with increasing mass $M_{\rm p}$ of the primary 
(i.e., the progenitor of the WD) \citep[see however][]{dm11}, and 
$\alpha_{\rm CE}\gtrsim 1$ for $1\lesssim M_{\rm p}/M_{\odot} \lesssim 2$, 
which seems to be in contrast with $\alpha_{\rm CE}\sim 0.25$
derived by \citet{rt12} from numerical simulations.
If there are planets/brown dwarfs around these
low-mass primaries, they can contribute extra orbital
energy to 
help expel the primary's envelope during the RGB/AGB 
phase, making the CE efficiency parameter within the canonical 
range $0 < \alpha_{\rm CE}< 1$.

\subsection{The Rebirth Periods of BMSPs}

Accretion onto the NS in I/LMXBs changes both the mass and spin
of the NS. The spin evolution of an accreting NS depends on
the interaction between the magnetosphere and the accretion
disk, as described as follows \citep[e.g.,][]{bv91}: the accretion disk is truncated
at the magnetospheric radius $R_{\rm m}$, which is 
close to the Alfv\'en radius of the disk
\[
R_{\rm m}\simeq R_{\rm A}=(\frac{B_{\rm s}^{2}R_{\rm s}^{6}}
{\dot{M}_{1}\sqrt{2GM_{1}}})^{2/7},
\]
 where $B_{\rm s}$ and $R_{\rm s}$ are
the surface magnetic field and the radius of the NS,
respectively. Material is channeled to the NS at the inner radius,
producing a spin-up torque. If the NS is spinning very rapidly, 
the disk may be truncated outside the corotation radius
$R_{\rm co}=(GM_1P^{2}/4\pi^{2})^{1/3}$, and the NS experiences
a centrifugal barrier that can inhibit
accretion \citep[i.e., the so-called ``propeller" effect;][]{is75}.
So the NS will eventually reach the 
equilibrium spin period such
that $ R_{\rm m}=R_{\rm co}$, or
\begin{equation}
P_{\rm eq} \simeq 2.4 B_{9}^{6/7}R_{6}^{18/7}M_{1}^{-5/7}
(\dot{M}_{1}/\dot{M}_{\rm Edd})^{-3/7}\,{\rm ms},
\end{equation}
where $B_{9}=B_{\rm s}/10^{9}$ G, and
$R_{6}=R_{\rm s}/10^{6}$ cm. This period can be regarded
as the beginning or rebirth period of recycled
pulsars when accretion terminates.

%A large fraction of BMSPs are found to be well below
%the rebirth line in the $B_{\rm s}-P_{\rm spin}$ 
%diagram given by Eq.~(10). 
The rebirth period of a MSP can be 
derived only when its actual spin-down
time is known. \citet{bk11} recently reported the optical discovery of the 
companion to the $\sim 2 M_{\odot}$  MSP J1614$-$2230. 
The optical colors show that the $0.5 M_{\odot}$  
companion is a 2.2 Gyr old CO WD. 
From the age of the WD, \citet{bk11} calculated the period of the 
pulsar at birth, and found that pulsar should be born with a spin close 
to its current value, implying
that the final accretion rate was $<10^{-2}\dot{M}_{\rm Edd}$. 
This value is two orders of magnitude smaller than the estimate from 
theoretical calculations by \citet{l11} and \citet{tlk11}. 
These authors  
suggested that the system
began as an IMXB consisting of a NS and a $\sim 4 M_{\odot}$ 
main-sequence secondary, which evolved to
be a CO WD with He envelope. The NS gained the most 
mass during the final LMXB phase lasting $\sim 5-10$ Myr
at near-Eddington rates.

We present a systematic view on the relation 
between $P_{\rm eq}/B_{9}^{6/7}$
and $P_{\rm orb}$ derived from theory and observations in Fig.~9. 
The solid lines outline the theoretically 
expected distribution of $P_{\rm eq}/B_{9}^{6/7}$
($\varpropto\dot{M}_{1}^{-3/7}M_{1}^{-5/7}$,  here $R_{6}$ is 
taken to be 1) from binary evolution calculations, and 
the symbols represent the observations of recycled pulsars 
\citep[data are taken from the ATNF Pulsar Catalogue;][]{m05}. 
Comparison between observations and theory shows
that, except in a few cases, 
the observational $P_{\rm spin}/B_{9}^{6/7}$ is generally
larger than $P_{\rm eq}/B_{9}^{6/7}$ by $\sim 1-2$ orders of 
magnitude, challenging the simple recycling theory\footnote{Here
we assume that the current spin periods of MSPs
are not far from their initial ones. This holds if their characteristic
ages are longer than the cooling 
ages of the WDs \citep[e.g.][]{l95} or the Hubble time.}.
Thus, some other mechanisms must work to
produce efficient spin down torque(s) or reduce the spin-up
torque due to mass accretion, if the stellar evolution models 
are correct. The proposed explanations can be 
summarized as follows.

(1) There is no definite spin equilibrium since GR 
may remove the angular momentum from 
the NS. This idea was first suggested by \citet{pp78} to account
for the cutoff in the spin distribution of NSs in LMXBs. 
The main emission mechanisms involve crustal mountains
\citep{b98}, magnetic deformations \citep{c02}, and
unstable $r$-modes \citep{a98} in the NSs. All these processes can produce
a substantial mass quadrupole moment and thus a
spin-down torque due to GR. The problem with 
this interpretation is that,
recent observational results on some millisecond
X-ray pulsars have shown that the efficiency of GR
induced spin-down might be too low to be responsible
for balancing the spin-up process during outbursts
\citep[see][and references therein]{hp11,phd12}. So the
following models focus on modifications of the equilibrium
period (Eq.~[10]) that the NS will finally reach.

(2) If the NS magnetic field lines can thread the
accretion disk, an extra magnetic torque will be
exerted on the NS \citep{gl79a}, and the equilibrium period will
be $\sim 1-3$ times that in Eq.~(10) \citep{gl79b,w95,lw96}.
\citet{a05} further suggested that the inner disk region 
may be geometrically thick
and sub-Keplerian, and dominated by radiation pressure,
if the mass transfer
rate is above a few percent Eddington accretion rate.
The coupling between the disk and magnetic field can reduce
the amount of angular momentum deposited
onto the NS from accretion by a factor $A$ (the
ratio of orbital angular velocity to Keplerian rotational
speed in the disk, $A<1$), compared with the
case of thin disks ($A = 1$), thus a decreased spin-up
torque results.
\citet{phd12} show that the existence of spin equilibrium 
as set by the disk$-$magnetosphere interaction is able to 
explain the observations of millisecond X-ray pulsars, if 
the spin-down torque coming from the interaction between 
the disk and field outside corotation is sufficiently large. 
However, the radial extent of the coupling between the 
star and the disk is highly controversial \citep[e.g.][]{mp05,g07}.
Stellar and disk winds may also take away the angular
momentum of the NS \citep{g97,g99,u06,r09,zl10}. In
addition, the abrupt torque reversals observed in 
disk-fed X-ray pulsars Her X-1, Cen X-3, and 4U 1626$-$67
\citep{b97} suggest that current models of 
disk$-$magnetosphere interaction
may have severe limitations.

(3) If Eq.~(10) does apply, a decaying mass transfer rate
at the end of the mass transfer process will result in relatively 
large rebirth period of BMSPs, provided that the $\dot{M}_2$ evolution 
time is longer than the NS spin evolution timescale \citep{j86,rst89}.  
More recently \citet{t12} suggested that 
this equilibrium will even be broken when 
$|\dot{M}_2|$ decreases substantially,
so that the magnetospheric radius $R_{\rm m}$ becomes
larger than the corotation radius $R_{\rm co}$, causing the NS 
to enter the propeller phase. 
A centrifugal barrier arises to expel matter entering 
the magnetosphere with a braking torque 
\begin{equation}
N\simeq \dot{M}_2R_{\rm m}^{2}\Omega_{\rm K}(R_{\rm m})
\end{equation}
(where $\Omega_{\rm K}(R_{\rm m})$ is the Keplerian 
angular velocity at $R_{\rm m}$) to act to slow down the 
NS. 
An alternatively
appropriate form for the accretion torque is \citep[e.g.][]{m99}
\begin{equation}
N\simeq -\dot{M}_2R_{\rm m}^{2}[\Omega_{\rm K}(R_{\rm m})
-\Omega_{\rm s}],
\end{equation}
where $\Omega_{\rm s}$ is the angular velocity of the NS.
With Eqs.~(11) and (12) we recalculate the spin evolution of the NS when the
donor star gradually decouples from its RL as in
\citet{t12}. In Fig.~10 the blue and red
lines denote the results calculated with Eqs.~(11) and (12), 
respectively\footnote{The discontinuity at the point where
the two lines meets is due to the fact that we use
Eq.~(12) to calculate the spin evolution even before the 
propeller phase in both cases.}. It is seen that the decaying mass transfer
rate indeed causes an increase in the spin period, which
is larger than the equilibrium period corresponding to 
the average mass transfer rate before decoupling. However,
The calculated discrepancy seems not to be large enough to explain the
deviation of the rebirth periods of BMSPs from that given by 
Eq.~(10), especially if we adopt the torque form of Eq.~(12).

(4) A decaying magnetic field may also change the rebirth
period of the BMSPs, if the field decay 
timescale is shorter than the spin evolution timescale, 
so that the final spin period deviates from the equilibrium 
period. During the decay of the NS magnetic field induced by accretion, 
the spin-up timescale
increases as $R_{\rm m}$ decreases because the accreting
matter carries less specific angular momentum, while the 
field decay timescale decreases  as $R_{\rm m}$ decreases
(field decay stops when $R_{\rm m}=R_{\rm s}$), resulting in a 
significant departure from the spin-up line \citep{b96,kb99}.

(5) The rebirth periods of BMSPs may be related to the long-term 
spin equilibrium with varying mass accretion rate. A
significant fraction of LMXBs are likely to be transient systems subject to
disk instability \citep{cfd12}. For example, an NS LMXB in a 10 hr
orbit would be transient if the average mass transfer rate is lower
than $\sim 10^{17}$ gs$^{-1}$. 
In addition, in narrow LMXBs, X-ray irradiation of the 
donor star could destabilize the mass transfer, and lead to mass transfer 
cycles \citep{h93} - mass transfer is spasmodic with phases 
of high mass transfer driven by the thermal expansion of the 
convective envelope of the irradiated donor alternating with
phases with low or no mass transfer, during  which the donor 
readjusts towards thermal equilibrium of the un-irradiated star.
The final spin period is thus determined by the 
spin-up during outbursts/high mass transfer phase 
balanced by the spin-down during 
quiescence/low mass transfer phase, which may be considerably 
longer than that 
attained via stable mass accretion \citep{l98,phd12}.

Currently our knowledge about the evolution of LMXBs 
and accretion disks is not sufficient to tell which one is the 
dominant factor in determining the rebirth periods of BMSPs.
Perhaps they are influenced by a combination of  
(at least some of) the above-mentioned mechanisms.

\subsection{Origin of Intermediate-Mass BMSPs with $P_{\rm orb} \lesssim$ 1 d}

It is well known that a large fraction of IMXBs will produce
binary pulsars with a CO WD companion.
As shown in \textbf{Fig.~3}, the orbital periods of these intermediate-mass
binary pulsars (IMBPs) are generally $>\sim 3-6$ d. This leaves the puzzle
that the binary pulsars with $P_{\rm orb} \lesssim$ 1 d and 
a CO WD companion
are hard to explain reasonably in any binary population \citep{d08}. 
In Table 1 we list the parameters of IMBPs 
with $P_{\rm orb} \lesssim$ 1 d. Except
PSR B0655$+$64, other pulsars share the characteristics of
rapid spin and low magnetic field.

The small orbital periods combined with massive WDs
suggest that this type of systems may have evolved
through a CE and spiral-in phase \citep{vt84}. 
Similar to that of the Double NSs, their direct progenitors could be 
binaries consisting of a He star and a NS star in a close and circular orbit,
as a result of spiral-in of a wide binary in late Case B or Case C 
mass transfer. The core of the companion was a He star which could 
already have gone through quite some He burning. After the spiral-in one 
then had a He star with already CO in its core.
During He-shell burning, the envelopes of the He stars, if their masses are
$ < 3.5\,M_{\sun}$, will expand and slowly transfer mass to the NS, 
which will have caused the spin-up and recycling of the NS. 
This interpretation seem to be
responsible for PSR B0655$+$64 \citep{vt84}, which has the longest
spin period (195.6 ms) and highest magnetic field
($B\sim 10^{10}$ G) among the pulsars in Table 1, 
implying that it has been partially
recycled with little mass accreted. 
For other pulsars, the
short periods and low magnetic fields ($B\sim 10^{8}$ G)
require that at least some $10^{-2}\, M_{\odot}$ material has been 
accreted by the NSs. Recently
\citet{clx11} presented the detailed binary evolution calculation 
for a binary consisting of a NS (of mass 1.3 $M_{\odot}$) and a
low-mass He star 
(of mass 1.0 $M_{\odot}$) with an initial orbital period of 0.5 d. 
They showed that the mass transfer seems to be able to 
spin up the NS's spin to milliseconds, producing BMSPs like PSR J1802$-$2124, 
which has a short orbital period $P_{\rm orb} \lesssim$ 1 d. 

Another possible way to form compact IMBPs 
invokes evolution of IMXBs with anomalous MB.
Traditionally MB is thought to work in low-mass,
main-sequence stars.
However, There are many intermediate-mass stars
which have anomalously high magnetic fields ($\gtrsim$ 100 G), i.e.,
Ap/Bp stars. In IMXBs, the irradiation-driven stellar wind from the
surface of the Ap/Bp donor star by the NSs  may couple with the 
high magnetic fields, resulting in angular momentum
loss at a rate \citep{jrp06},
\begin{equation}
\dot{J}_{\rm MB} = -B_2(\dfrac{\psi\dot{M}_{2}M_1}{a^{3}})^{1/2}
(\frac{R_2^{15}}{GM_{2}^{3}})^{1/4},
\end{equation}
where $B_2$ is the magnetic field strength at the surface
of the donor star, and $\psi$ is a parameter with typical value 
$\psi/c^{2}\sim 10^{-6}$,
combining the wind-driving energy conversion efficiency and
irradiation geometry. \citet{jrp06} explored the effect 
of anomalous MB in the formation 
of compact BH LMXBs, in which the orbital energy
of the companion star, if of low-mass initially, would be 
unable to expel the envelope of the BH progenitor.  
For the black hole LMXB XTE J1118$+$480, \citet{g12} measured a
decay rate of the orbital period 
$\dot{P}_{\rm orb}=-1.83\pm0.66 $ ms\,yr$^{-1}$.
This value is much larger than predicted by 
most conventional MB and mass loss models 
for the binary parameters of XTE J1118$+$480, unless 
the companion star has a surface magnetic field 
$B_2 \geq 10-20$ kG to enhance MB.

In Fig.~11 we show the evolution of a binary initially 
consisting of a 1.8 $M_{\odot}$ NS and a 3.5 $M_{\odot}$ donor 
with $P_{\rm orb}\simeq 2.5$ d. We assume that the
donor possesses a surface magnetic field $B_2= 500$ G.
RLOF starts at the age $\sim2.4\times10^{8}$ yr
with a mass transfer rate $\sim10^{-5} M_{\odot}$\,yr$^{-1}$. 
Because the donor is initially more massive than the NS, 
the orbital period decreases rapidly. Most of the transferred 
mass is assumed to be ejected out of the binary
in the vicinity of the NS, taking away both mass and orbital
angular momentum. The orbit starts to widen when
$M_2\sim 2\,M_{\odot}$.
When the donor mass decreases to be $\sim 0.8 M_{\odot}$, 
the mass transfer rate reduces to be
$< 10^{-7}M_{\odot}$\,yr$^{-1}$. In the subsequent 
$\sim8\times10^{7}$ yr, MB begins to dominate the evolution. 
The orbital period keeps decreasing,  the NS accretes 
most mass in this phase and is spun up to milliseconds. 
The final outcome is a tight binary system
consisting  of a $\sim 2.1\,M_{\odot}$ NS and 
a $\sim 0.5 M_{\odot}$ WD with $P_{\rm orb}\sim$ 0.4 d.

The anomalous MB model will not work if the 
NS is initially of low-mass ($\sim 1\,M_{\odot}$), since the 
allowed parameter space for forming binary pulsars will be
very small (see Fig.~1). 
Additionally we note that models involving
accretion-induced collapse of an ONeMg WD 
cannot account for the $P_{\rm orb} $ distribution
of these BMSPs \citep{sl00}. 

\section{Summary}

This work is motivated by the fact the evolution of I/LMXBs with
canonical NSs seems to meet difficulties in explaining some of the 
observational characteristics of BMSPs, and the
measurements of the NS masses indicate a
wide distribution $\sim 1-1.8\,M_{\odot}$.
We have preformed numerical
calculations of the evolution of I/LMXBs
consisting of a 1.0 or $1.8 M_{\odot}$ NS and a 
$1.0-6.0 M_{\odot}$ donor star, to investigate its dependence
on the initial NS mass, and the properties of the descendent
BMSPs.
The main results can be summarized as follows.

1. The allowed parameter space in the initial $P_{\rm orb}
-M_2$ diagram for forming recycled pulsars 
increases with increasing NS mass. This 
may help explain the formation of BMSPs with 
$P_{\rm orb}\gtrsim 60$ d  and their distribution in the
 $P_{\rm orb}-M_{\rm WD}$ diagram.
Alternatively, some of these wide binary pulsars may be formed 
through mass transfer driven by planet/brown dwarf-involved 
CE evolution.

2. The equilibrium spin periods $P_{\rm eq}$ of accreting
NSs in LMXBs derived from the standard 
magnetosphere-accretion disk interaction model are in general
shorter than the observed spin periods  of BMSPs
by more than one order of magnitude. This implies that either the simple
equilibrium spin model does not apply for the spin evolution
in accreting NSs, or there are
other mechanisms/processes to spin down the NSs when forming
BMSPs.

3. Some of the compact IMBPs
might have evolved from IMXBs in which the companion
star were strongly magnetized Ap/Bp stars with enhanced MB.

4. Our calculations doubt the suggestion that the orbital 
period gap ($\sim20 -60$ d) of BMSPs is related to the 
occurrence of the bump-related detached phase in the 
LMXB evolution, since  the accretion disk would become
unstable at earlier time.

\begin{acknowledgements}

We are grateful to Ed van den Heuvel for helpful discussions and
the referee for constructive comments.
This work was supported by the Natural Science Foundation of China
under grant number 11133001, the
National Basic Research Program of China (973 Program 2009CB824800),
and the Qinglan project of Jiangsu Province.

\end{acknowledgements}

\clearpage

\begin{figure}

\plottwo{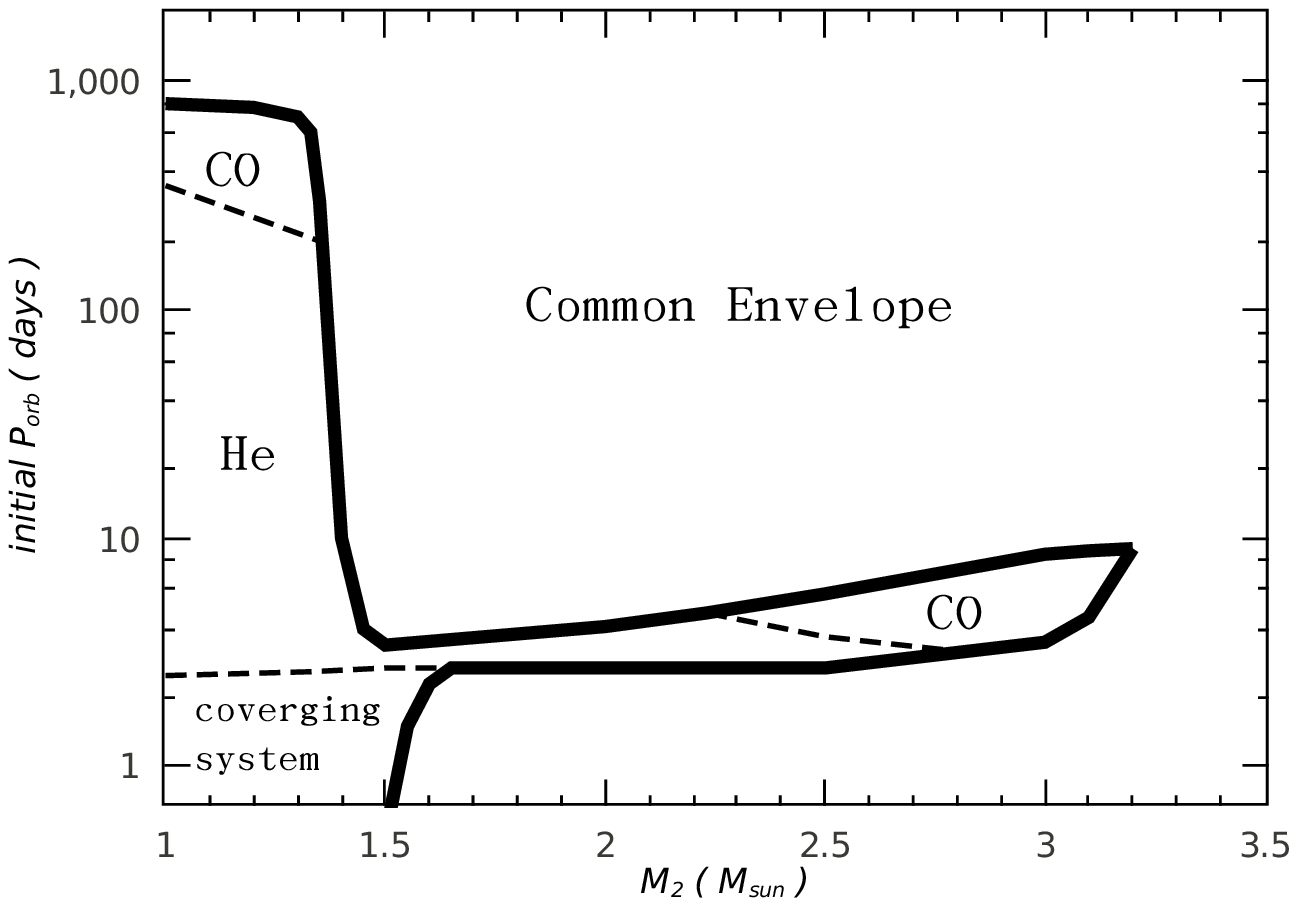}{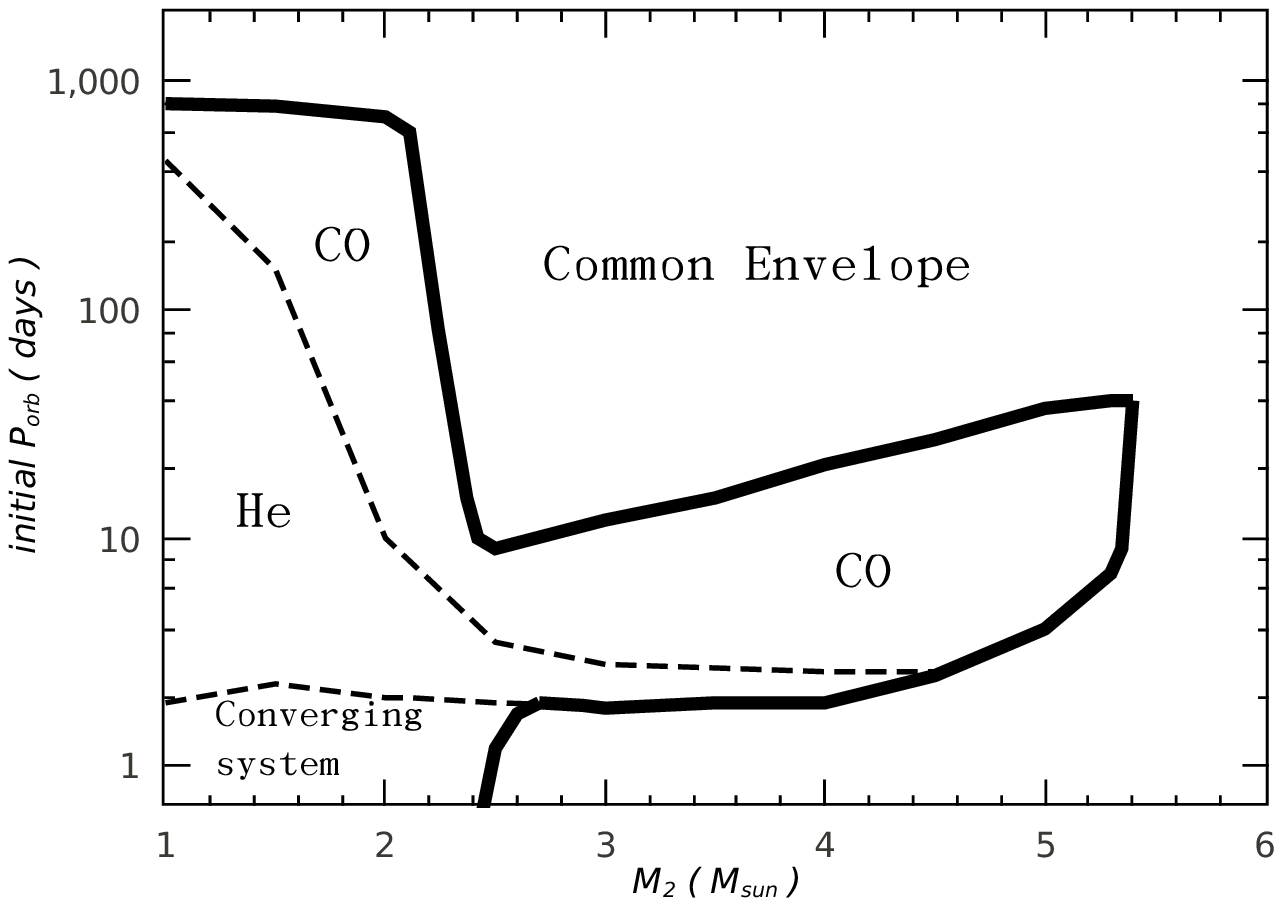}
\caption{The thick lines confine the allowed parameter space 
in the initial orbital period vs. donor mass plane for I/LMXBs
to successfully form binary pulsars with He and CO WD
companions. In the left and right panels the initial mass of the NS is taken
to be $1.0\,M_{\odot}$ and $1.8\,M_{\odot}$, respectively.     \label{figure1}}

\end{figure}

\begin{figure}[h,t]
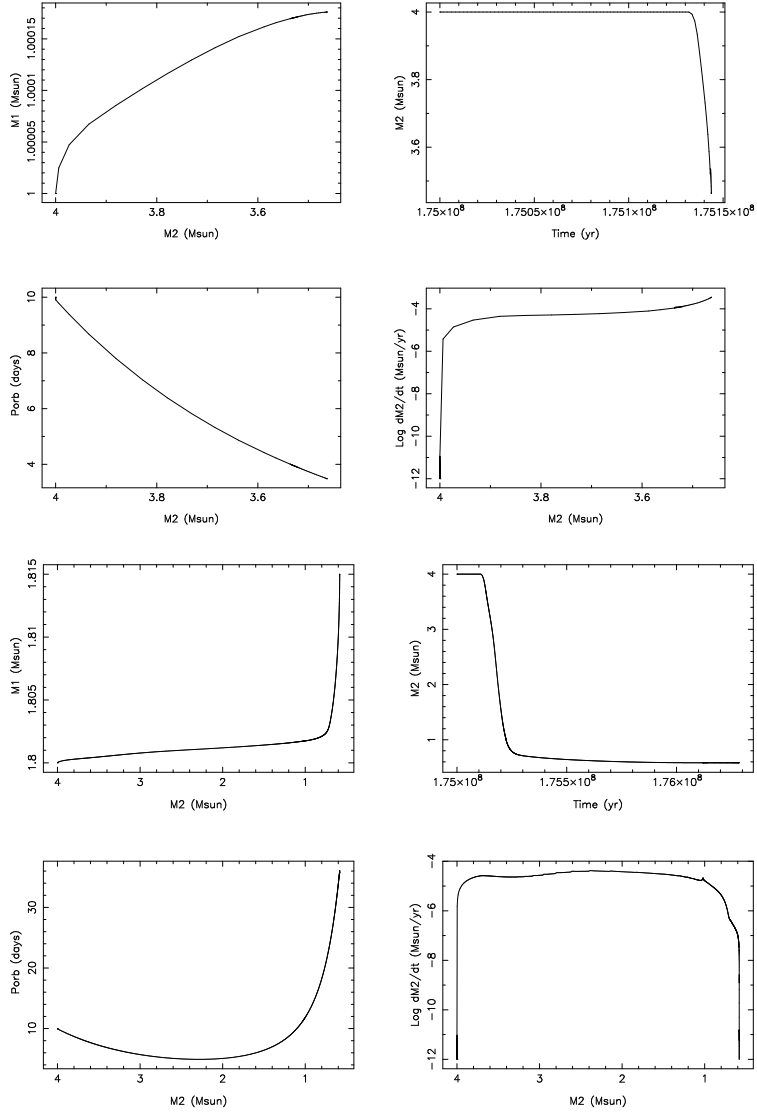


\centerline{\includegraphics[angle=-90,width=0.6\textwidth]{f2a.eps}}
\bigskip
\centerline{\includegraphics[angle=-90,width=0.6\textwidth]{f2b.eps}}

\caption{Evolution of the  NS mass $M_{1}$, the donor mass $M_{2}$,
the obital period $P_{\rm orb}$, and the mass transfer rate $\dot{M}_{2}$ 
for a IMXB consisting of a 1.0 $M_{\odot}$
(upper tow panels)  or 1.8 $M_{\odot}$ (lower two panels) NS.     \label{figure2}}

\end{figure}

\begin{figure}[h,t]

\plottwo{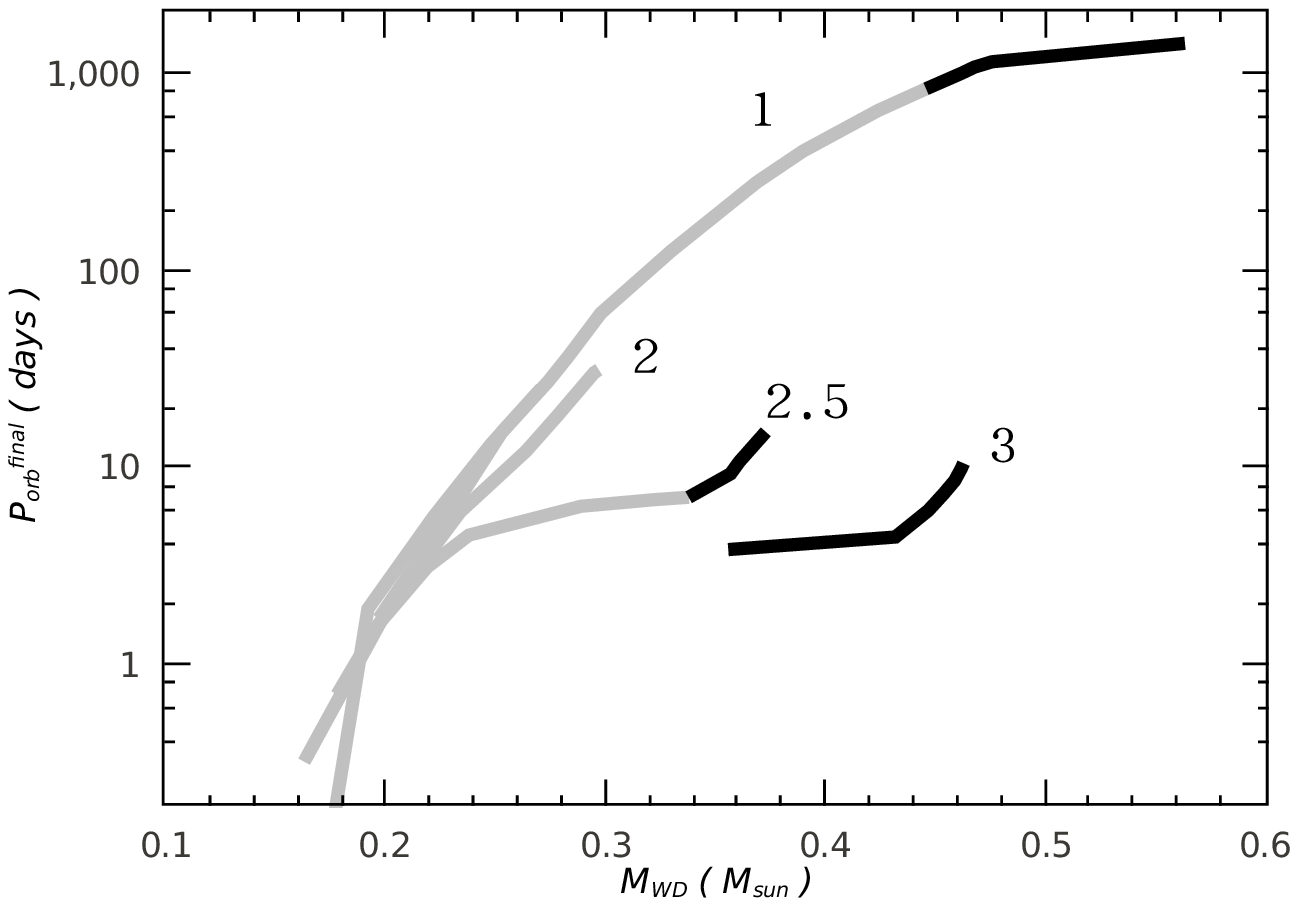}{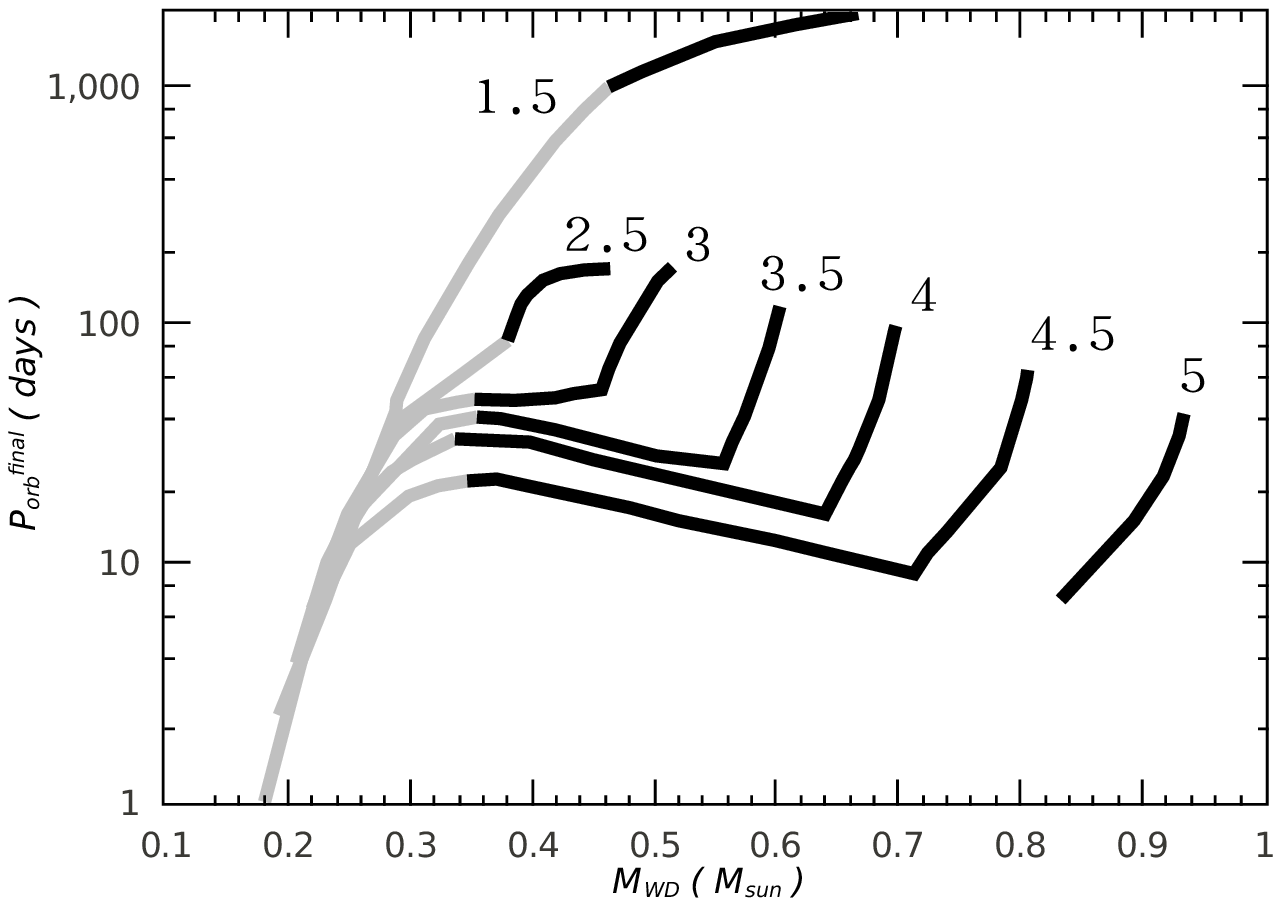}
%\centerline{\includegraphics[angle=0,width=0.40\textwidth]{f4.eps}}
\caption{The final orbital period $P_{\rm orb}$ as a function of 
the WD mass $M_{\rm WD}$ for binary pulsars 
with an initial $1.0\,M_{\odot}$ (left) or $1.8\,M_{\odot}$ (right) NS.
The number next to each curve
denotes the initial mass of the donor star (the progenitor of
the WD). The gray and black curves
represent binary pulsars with He and CO WD companions, 
respectively.\label{figure3}}

\end{figure}

\begin{figure}[h,t]
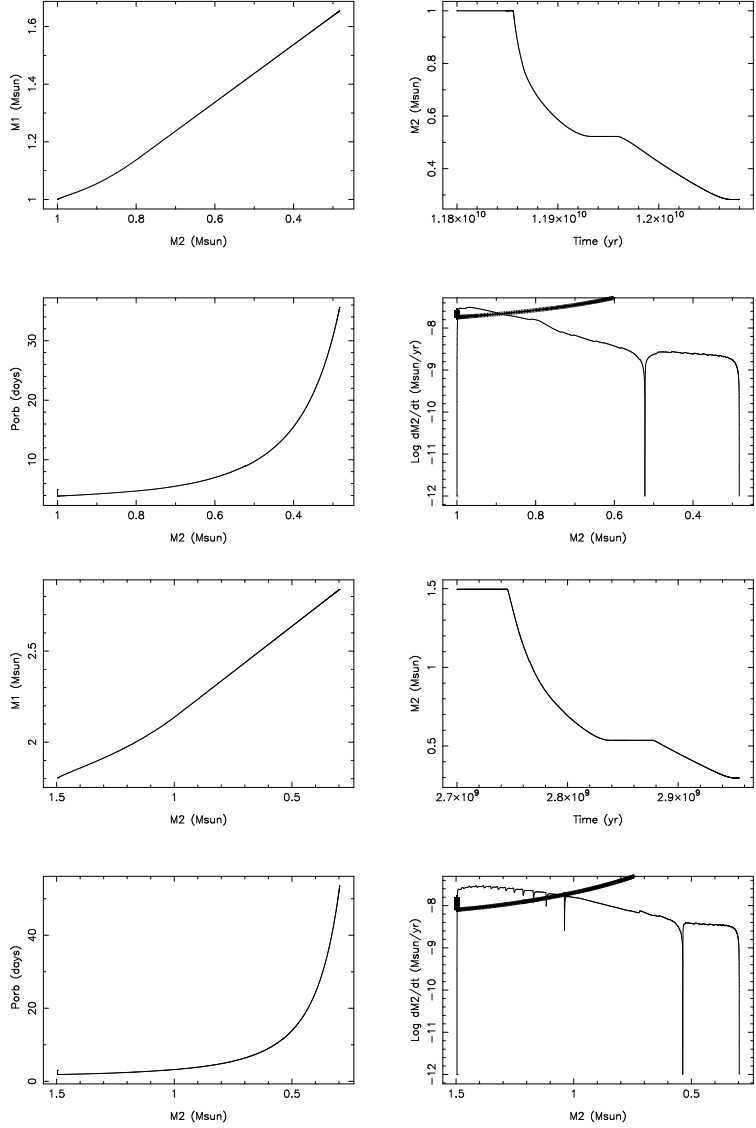


\centerline{\includegraphics[angle=-90,width=0.6\textwidth]{f4a.eps}}
\bigskip
\centerline{\includegraphics[angle=-90,width=0.6\textwidth]{f4b.eps}}
\caption{Evolution of the NS mass  $M_{1}$, the donor mass
$M_{2}$, the obital period $P_{\rm orb}$,
and the mass transfer rate for a LMXB with a 1.0 $M_{\odot}$ NS 
and  a 1.0 $M_{\odot}$ initial donor (upper two panels)
or with a 1.8 $M_{\odot}$ NS and a 1.5 $M_{\odot}$ initial donor (lower 
two panels). The
thick lines represent the critical mass
transfer rate for unstable accretion disks.  \label{figure4}}

\end{figure}

%\begin{figure}[h,t]

%\centerline{\includegraphics[angle=-90,width=1.0\textwidth]{f5.eps}}
%\caption{Evolution of the NS mass  $M_{1}$, the donor mass
%$M_{2}$, the obital period $P_{\rm orb}$,
%and the mass transfer rate for a LMXB with a 1.8 $M_{\odot}$ NS. The
%thick curve correspond to the critial mass
%transfer rate for unstable accretion disk.   \label{figure5}}

%\end{figure}

\begin{figure}[h,t]
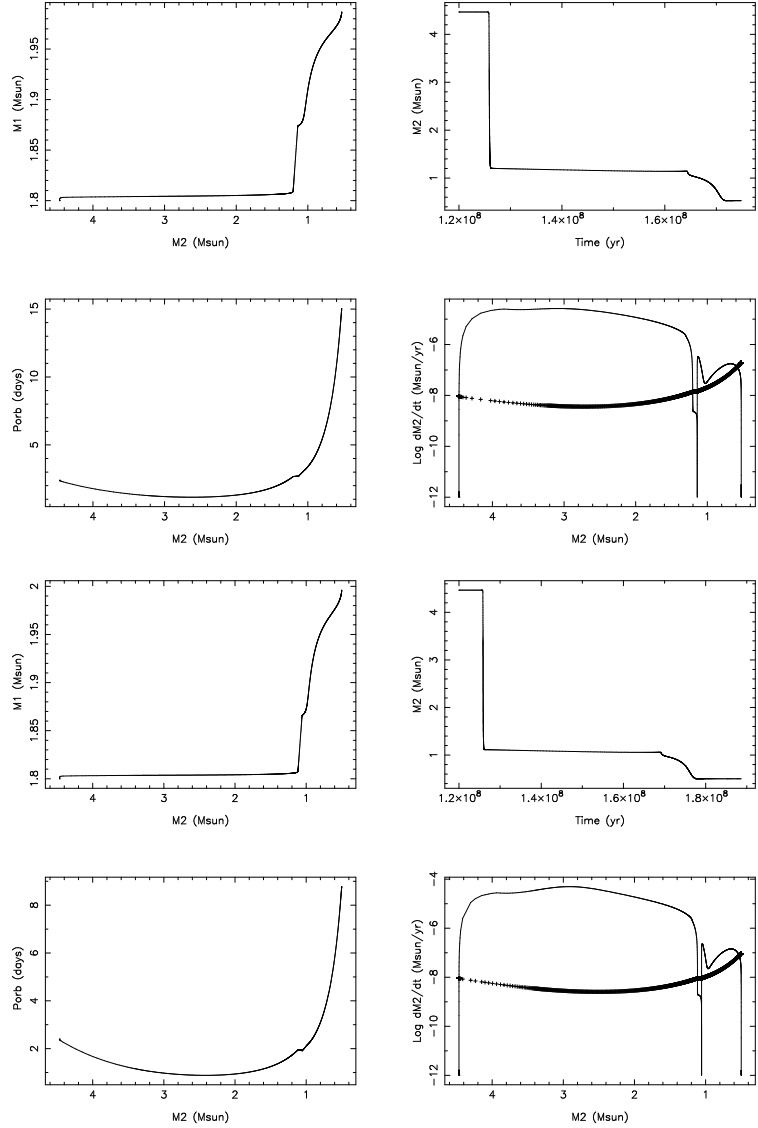


\centerline{\includegraphics[angle=-90,width=0.6\textwidth]{f5a.eps}}
\bigskip
\centerline{\includegraphics[angle=-90,width=0.6\textwidth]{f5b.eps}}
\caption{This figure shows the effect of outflow from the $L_2$ 
point on the formation of MSP J1614$-$2230. The upper two and
lower two panels correspond to $\delta = 0$ and 0.04, respectively.
\label{figure5}}

\end{figure}

\begin{figure}[h,t]

\plottwo{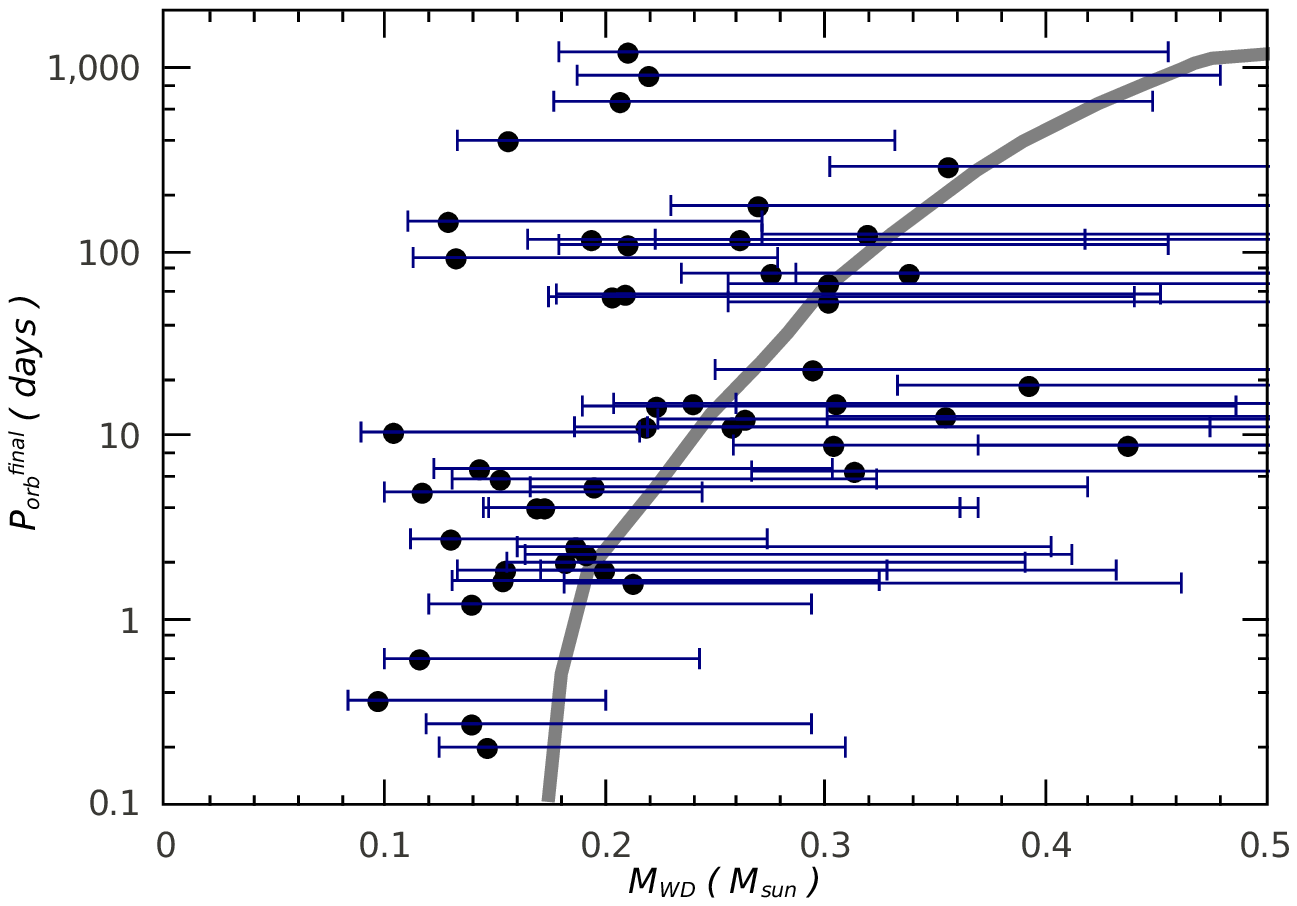}{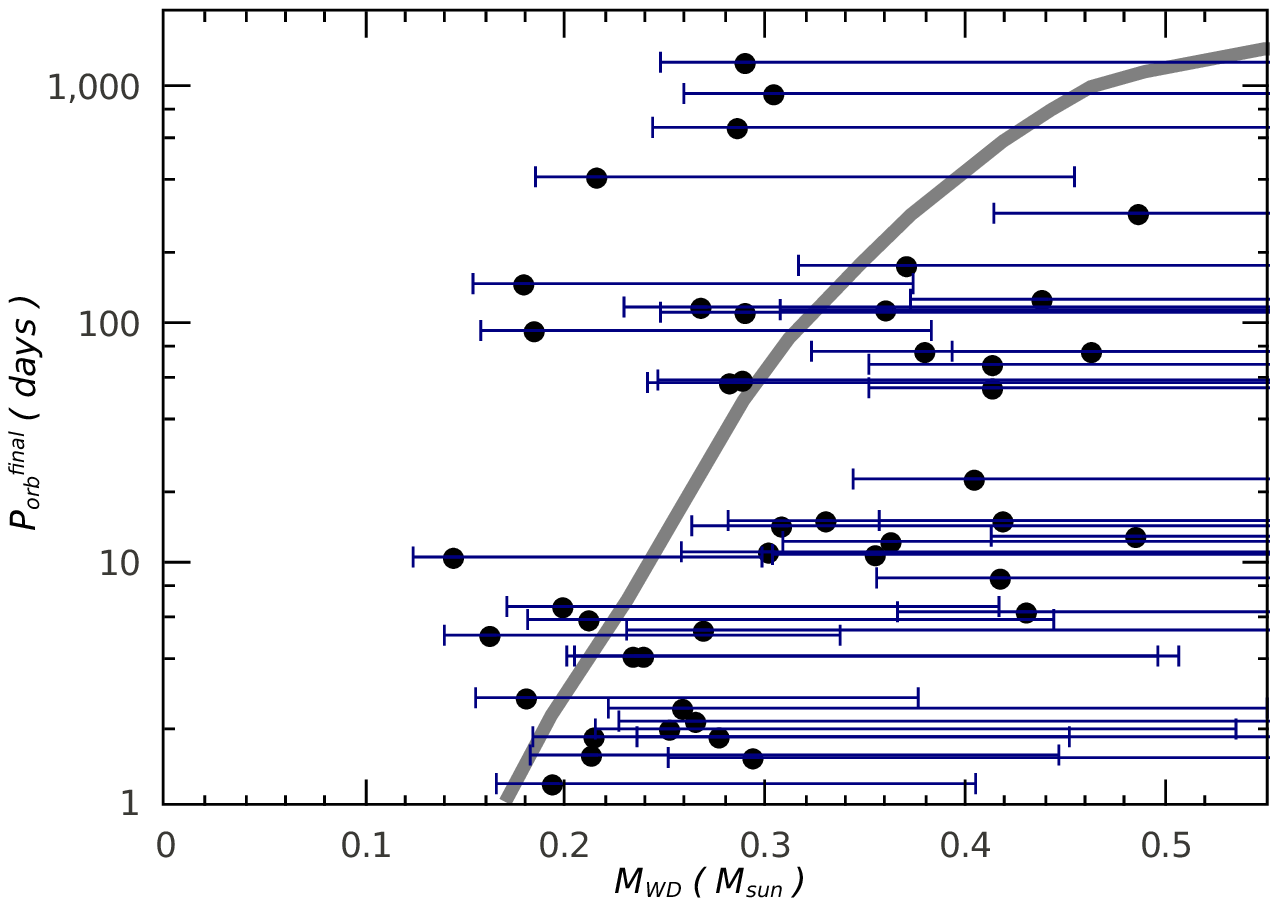}
%\centerline{\includegraphics[angle=0,width=0.40\textwidth]{f6.eps}}
\caption{The orbital period as a function of the WD mass for 
MSPs with possible He WD companions. In the left and right panels
the thick lines are theoretically expected relations for
NSs with an initial mass of 1.0 $M_{\odot}$  and
1.8 $M_{\odot}$, and the dots correspond to 
binary pulsars with an assumed mass of 1.2 $M_{\odot}$ and 2.0 $M_{\odot}$,
respectively. The error bars of the WD masses  
cover the 90\% probability mass range for randomly
oriented orbits having $i = 90^{\circ}$ to $i=26^{\circ}$.
\label{figure6}}

\end{figure}

\begin{figure}[h,t]

\plottwo{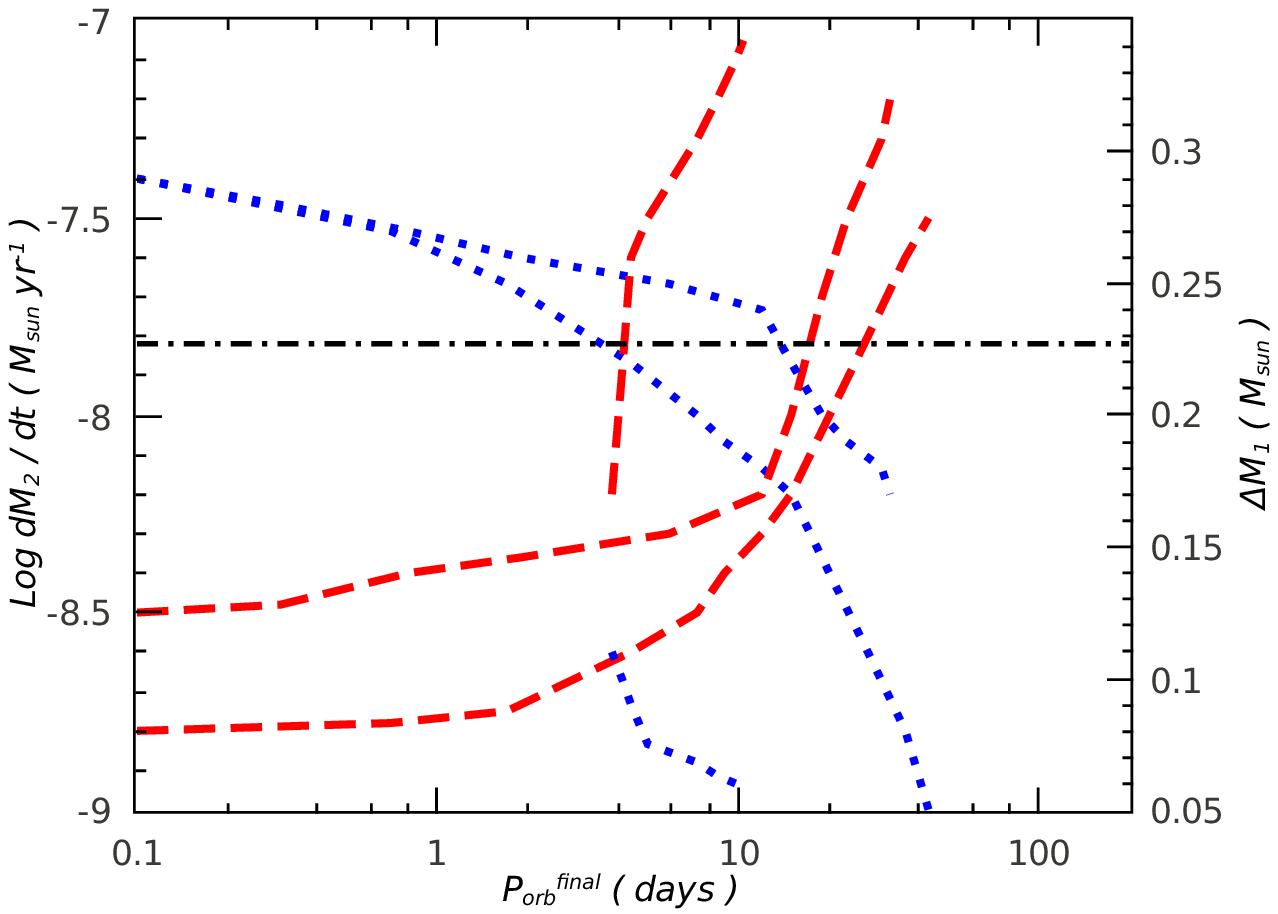}{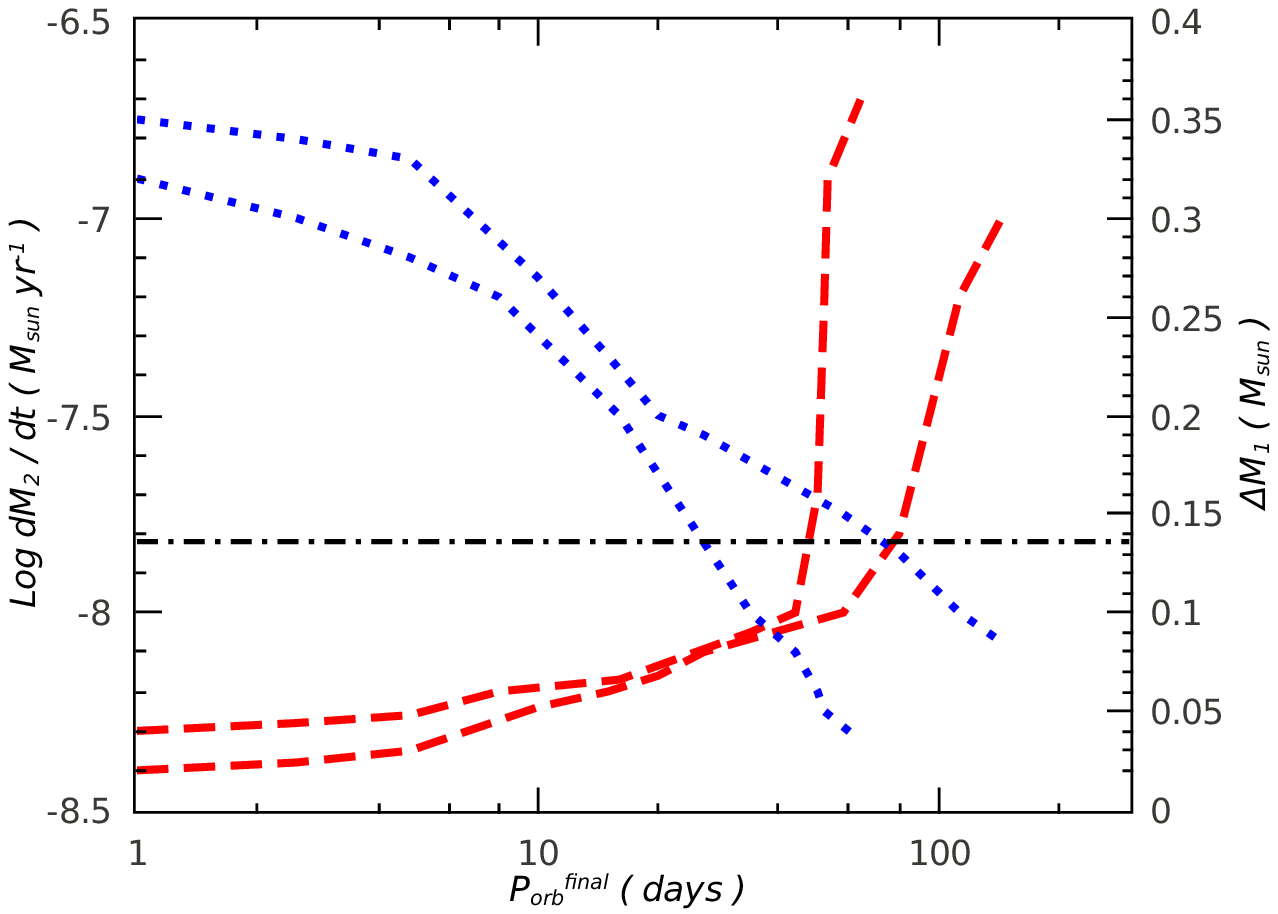}
%\centerline{\includegraphics[angle=0,width=0.40\textwidth]{f7.eps}}
\caption{The mass transferred rate (red dashed curves)
and accreted masses (blue dotted curves)
of the NS as a function of the final
orbital period for binaries with a 1.0 $M_{\odot}$ (left panel) or
1.8 $M_{\odot}$ (right panel) NS.
The black dot-dashed line show the critical accretion rate
$\dot{M}_{\rm Edd}$. In the left panel, the red dashed (blue dotted) curves,
from bottom (top) to top (bottom), correspond to
1.0, 2.0, and 3.0 $M_{\odot}$ donor stars, respectively. In the
right panel, plotted are those in the cases of 2.0 and 3.0 $M_{\odot}$ 
donor stars. \label{figure7}}

\end{figure}

\begin{figure}[h,t]

\centerline{\includegraphics[angle=0,width=0.40\textwidth]{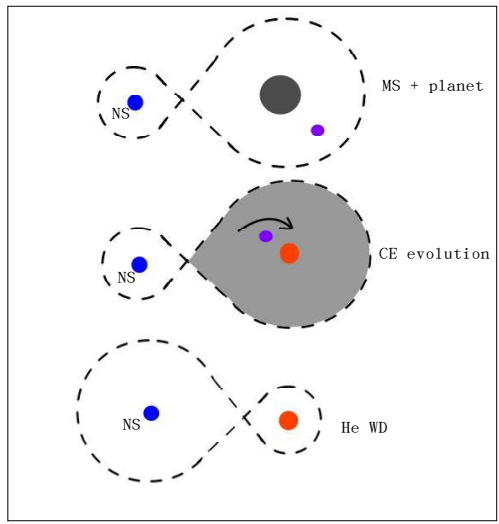}}
\caption{This schematic plot shows the possible formation process of
wide BMSPs with a He WD companion. We assume that in a wide
binary (with orbital period at least longer than tens of days) the 
secondary star initially possesses a (or multiple) substellar object(s),
which spirals into the stellar envelope when it expands and 
develops a He core. The CE evolution may also cause the secondary to fill 
its Roche-lobe, and transfer mass rapidly to the NS. After the envelope
is expelled, a wide binary with a NS and a He WD is produced. 
If the NS can accrete sufficient mass from its companion, it will appear
as a MSP. 
\label{figure8}}

\end{figure}

%
%\begin{figure}[h,t]

%\centerline{\includegraphics[angle=0,width=0.40\textwidth]{f9.eps}}
%\caption{The mass transfered rate ( red dashed curves )
%and accreted masses ( blue dashed curves )
%of the NS as a function of the final
%obital period for binaries with initial 1.8 $M_{\odot}$
%NS. The black dashed line show the position of
%$\dot{M}_{edd}$ as contrast, and the red dashed curves,
%from bottom to top, correspnod to
%2.0, 3.0 $M_{\odot}$ donor star, respectively. \label{figure9}}

%\end{figure}

\begin{figure}[h,t]

\centerline{\includegraphics[angle=0,width=0.50\textwidth]{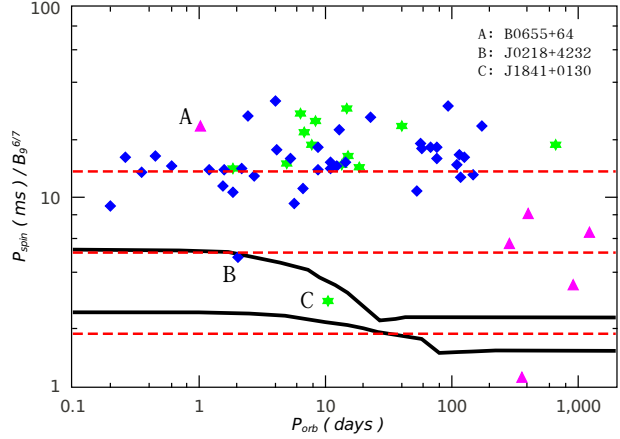}}
\caption{The black lines show the relationship between 
$P_{\rm eq}/B_{9}^{6/7}$
and $P_{\rm orb}$ for all systems from our calculations.
Blue diamonds, green stars, and magenta triangles
represent binary pulsars with $P_{\rm spin} \leq 10$ ms, 10 ms
$<P_{\rm spin} \leq$ 100 ms, and $P_{\rm spin} >$
100 ms, respectively. The red dashed lines from top
to bottom correspond to mass accretion rate of  0.01,
0.1, and $1.0 \dot{M}_{\rm Edd}$ for a $1.4 M_{\odot}$ NS, 
respectively. \label{figure9}}

\end{figure}

\begin{figure}[h,t]

\centerline{\includegraphics[angle=0,width=0.50\textwidth]{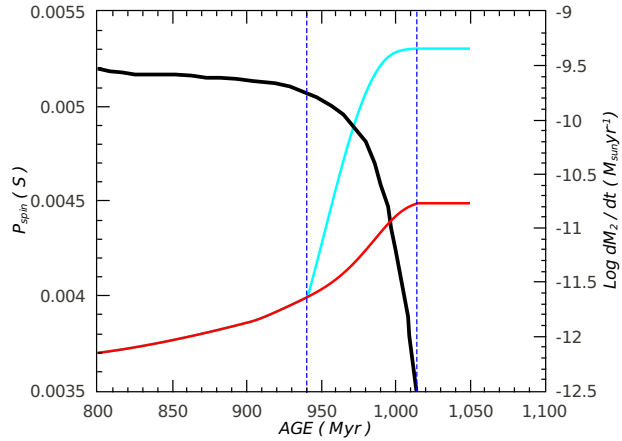}}
\caption{Spin evolution of an accreting NS at the end of
mass transfer phase as in \citet{t12}. 
The black line denotes the evolution of the 
mass transfer rate, and the blue and red lines denote the spin
evolution governed by Eqs.~(11) and (12), respectively. The
two vertical dashed lines confine the propeller phase.
\label{figure10}}

\end{figure}

\begin{figure}[h,t]

\centerline{\includegraphics[angle=-90,width=0.6\textwidth]{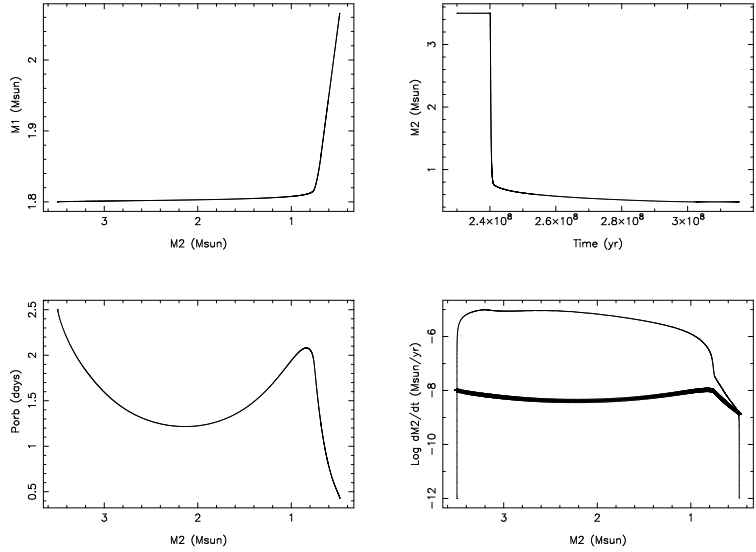}}
\caption{The evolution of the NS mass, the donor mass, 
the orbital period, and the mass transfer rate for a 
1.8 $M_{\odot}$ NS IMXB. 
With anomalous MB, the final outcome is a IMBP with 
orbital less than 1 d.
The thick line corresponds to the critical mass
transfer rate for unstable accretion disks.  
\label{figure11}}

\end{figure}

\clearpage
\begin{table}
\begin{center}
\caption{Parameters of Intermediate-Mass Binary Pulsars with $P_{\rm orb}\lesssim 1$ d$^a$ \label{tbl-1}}
\begin{tabular}{lcccc}
\tableline\tableline
Pulsar & $P_{\rm spin}$ (ms) & $P_{\rm orb}$ (d) & $M_{\rm WD}^b$ & $B$ (G) \\
\tableline
B0655$+$64     &195.6 & 1.0287 & 0.80 & $1.17\times 10^{10}$ \\
J1435$-$6100  & 9.3     & 1.3549 & 1.08 & $4.84\times 10^{8}$  \\
%J1748$-$2446N & 8.7   &  0.3855 & 0.56 &      \\
J1757$-$5322    & 8.9   &  0.4533 & 0.67 & $4.89\times 10^8$  \\
J1802$-$2124    &12.6  &  0.6989 & 0.98 & $ 9.69\times 10^8$  \\
J1952$+$2630   & 20.7 &  0.3919 & 1.13 &       \\
\tableline
\end{tabular}
\tablenotetext{a}{Data are taken from the ATNF pulsar catalogue.}
\tablenotetext{b}{Median mass for the inclination angle $i=60\degr$.}
\end{center}
\end{table}


\begin{thebibliography}{28}
\expandafter\ifx\csname natexlab\endcsname\relax\def\natexlab#1{#1}\fi

\bibitem[{{Andersson}(1998)}]{a98}
Andersson, N. 1998, \apj, 502, 708

\bibitem[{{Andersson} et~al.(2005)}]{a05}
Andersson, N., Glampedakis, K., Haskell, B., \& Watts, A. L.
2005, \mnras, 361, 1153

\bibitem[{{Bailyn} et~al.(1989)}]{b89}
Bailyn, C. D., Garcia, M. R., \& Grindlay, J. E. 1989, \apj, 344, 786

\bibitem[{{Barziv} et~al.(2001)}]{b01}
Barziv, O., Kaper, L., Van Kerkwijk, M. H., Telting, J. H., \& van Paradijs, J.
2001, \aap, 377, 925

\bibitem[{{Bear \& Soke}(2010)}]{bs10}
Bear, E. \& Soker, N. 2010, New Astronomy, 15, 483

\bibitem[{{Bhalerao \& Kulkarni}(2011)}]{bk11}
Bhalerao, V. B. \& Kulkarni, S. R. 2011, \apj, 737, L1

\bibitem[{{Bhattacharya \& van den Heuvel}(1991)}]{bv91}
Bhattacharya, D. \& van den Heuvel, E. P. J. 1991, Phys. Rev., 203, 1

\bibitem[{{Bildsten}(1998)}]{b98}
Bildsten, L. 1998, \apj, 501, L89

\bibitem[{{Bildsten} et~al.(1997)}]{b97}
Bildsten, L., Chakrabarty, D., Chiu, J., Finger, M. H., \&  Koh, D. T.
et al. 1997, \apjs, 113, 367

\bibitem[{{Burderi} et~al.(2002)}]{bdb02}
Burderi, L., D'Antona, F., \& Burgay, M. 2002, \apj, 574, 325

\bibitem[{{Burderi} et~al.(1996)}]{b96}
Burderi, L., King, A. R., \& Wynn, G. A. 1996, \mnras, 283, L63

\bibitem[{{Cassan} et~al.(2012)}]{c12}
Cassan, A., Kubas, D., Beaulieu, J.-P., Dominik, M., Horne, K.
et al. 2012, \nat, 481, 167

\bibitem[{{Chen} et~al.(2011)}]{clx11}
Chen, W.-C., Li, X.-D., \& Xu, R.-X., 2011, \aap, 530, 104

\bibitem[{{Coriat} et~al.(2012)}]{cfd12}
Coriat, M., Fender, R. P., \& Dubus, G. 2012, arXiv:1205.5038

\bibitem[{{Cutler} (2002)}]{c02}
Cutler C., 2002, Phys. Rev. D, 66, 084025

\bibitem[{{D'Antona} et~al.(2006)}]{dvl06}
D'Antona F., Ventura P., Burderi L.,  Di Salvo T., \& Lavagetto G. ,
et al. 2006, \apj, 640, 950

\bibitem[{{Davis} et~al.(2012)}]{dkk12}
Davis, P. J., Kolb, U., \& Knigge, C. 2012, \mnras, 419, 287

\bibitem[{{Deloye}(2008)}]{d08}
Deloye C. J. 2008, in 40 YEARS OF PULSARS: Millisecond Pulsars, Magnetars and More. AIP Conference Proceedings, Eds. C. Bassa, Z. Wang, A. Cumming, \& V. M. Kaspi, Vol. 983, p. 501

\bibitem[{{De Marco} et~al.(2011)}]{dm11}
De Marco, O., Passy, J., Moe, M., Herwig, F., \& Mac Low, M. et al. 2011,
\mnras, 411, 2277

\bibitem[{{Demorest} (2010)}]{d10}
Demorest P. B., Pennucci T., Ransom S. M., Roberts
M. S. E., \& Hessels J. W. T., 2010, \nat, 467, 1081

\bibitem[{{De Vito \& Benvenuto}(2010)}]{db10}
De Vito, M. A. \& Benvenuto, O. G., 2010, \mnras, 401, 2552

\bibitem[{{Dubus} et~al.(1999)}]{dlhc99}
Dubus, G., Lasota, J.-P., Hameury, J.-M., \& Charles, P. 1999, \mnras, 303, 139

\bibitem[{{Eggleton}(1971)}]{e71}
Eggleton, P. P. 1971, \mnras, 151, 351

\bibitem[{{Eggleton}(1972)}]{e72}
Eggleton, P. P. 1972, \mnras, 156, 361

\bibitem[{{Eggleton}(1983)}]{e83}
Eggleton, P. P. 1983, \apj, 268, 368

\bibitem[{{Ergma} et~al.(1998)}]{e98}
Ergma, E., Sarna, M. J., \& Antipova, J. 1998, \mnras, 300, 352

\bibitem[{{Faulkner}(1971)}]{f71}
Faulkner, J. 1971, \apj, 170, L99

\bibitem[{{Ferdman} et~al.(2010)}]{f10}
Ferdman, R. D., Stairs, I. H., Kramer, M., McLaughlin, M. A., 
\& Lorimer, D. R. et al. 2010, \apj, 711, 764

\bibitem[{{Fu \& Li}(2011)}]{fl11}
Fu, L. \& Li, X.-D. 2011, RAA, 11, 1457

\bibitem[{{Ghosh}(2007)}]{g07}
Ghosh, P. 2007,  Rotation and Accretion Powered Pulsars (World Scientific,
London), p. 772

\bibitem[{{Ghosh \& Lamb}(1979a)}]{gl79a}
Ghosh, P. \& Lamb, F. K. 1979a, \apj, 232, 259

\bibitem[{{Ghosh \& Lamb}(1979b)}]{gl79b}
Ghosh, P. \& Lamb, F. K. 1979b, \apj, 234, 296

\bibitem[{{Gonz\'alez Hern\'andez} et~al.(2012)}]{g12}
Gonz\'alez Hern\'andez, J.,  Rafael Rebolo, R. \&  
Casares, J. 2012, \apj, 744L, 25

\bibitem[{{Goodson} et~al.(1999)}]{g99}
Goodson, A. P., B\"ohm, K. H., \& Winglee, R. M. 1999, \apj, 524, 142

\bibitem[{{Goodson} et~al.(1997)}]{g97}
Goodson, A. P., Winglee, R. M., \& B\"ohm, K. H. 1997, \apj, 489, 199

\bibitem[{{Hameury} et~al.(1993)}]{h93}
Hameury, J. M., King, A. R., Lasota, J. P., \& Raison, F. 1993, \aap, 277, 81

\bibitem[{{Han} et~al.(1994)}]{h94}
Han, Z., Podsiadlowski, P., \& Eggleton, P.P, 1994, \mnras, 270, 121

\bibitem[{{Haskell \& Patruno}(2011)}]{hp11}
Haskell, B. \& Patruno, A. 2011, \apj, 738, 14

\bibitem[{{Heger} et~al.(2003)}]{h03}
Heger, A., Fryer, C. L., Woosley, S. E., Langer, N. \& Hartmann, D. H.
2003, \apj, 591, 288

\bibitem[{Iben \& Livio(1993)}]{il93}
Iben, Jr. \& Livio, M., 1993, \pasp, 105, 1373

\bibitem[{Jeffrey(1986)}]{j86}
Jeffrey, L. C. 1986, \nat, 319, 384

\bibitem[{{Illarionov \& Sunyaev}(1975)}]{is75}
Illarionov, A. F. \& Sunyaev, R. A. 1975, \aap, 39, 185

\bibitem[{{Joss} et~al.(1987)}]{j87}
Joss, P. C., Rappaport, S. A., \& Lewis, W. 1987, \apj, 319, 180

\bibitem[{{Justham} et~al.(2006)}]{jrp06}
Justham, S., Rappaport, S., \& Podsiadlowski, P.
2006, \mnras, 366, 1415

\bibitem[{{King}(1988)}]{k88}
King, A. R. 1988, QJRAS, 29, 1

%\bibitem[{{King \& Kolb }(1995)}]{kk95}
%King, A. R. \& Kolb, U., 1995, \apj, 439, 330

\bibitem[{{King} et~al.(1997)}]{k97}
King, A. R., Kolb, U., \& Sienkiewicz, E. 1997, \apj, 488, 89

\bibitem[{{King} et~al.(2003)}]{k03}
King A. R., Rolfe D. J., Kolb U., \& Sshenker K., 2003, \mnras, 341, L35

%\bibitem[{{Kulkarni \& Narayan}(1988)}]{kn88}
%Kulkarni, S. R. \& Narayan, R. 1988, \apj, 335, 755

\bibitem[{{Kiziltan, Kottas, \& Thorsett}(2010)}]{k10}
Kiziltan, B., Kottas, A., \& Thorsett, S. E. 2010, arXiv:1011.4291

\bibitem[{{Koester \& Reimers}(2000)}]{kr00}
Koester, D. \& Reimers, D. 2000, \aap, 364, L66

\bibitem[{{Kolb} et~al.(2000)}]{kdkr00}
Kolb, U., Davies, M. B., King, A., \& Ritter, H. 2000, \mnras, 317, 438

\bibitem[{{Konar  \& Bhattacharya}(1999)}]{kb99}
Konar, S. \& Bhattacharya, D. 1999, \mnras, 303, 588

\bibitem[{{Landau \& Lifshitz}(1959)}]{ll59}
Landau, L. D. \& Lifshitz, E. M. 1959, The Classical Theory of Fields
(Oxford: Pergamon Press)

\bibitem[{{Lasota}(2001)}]{l01}
Lasota, J.-P. 2001, New Astron. Rev., 45, 449

\bibitem[{{Li} et~al.(1998)}]{l98}
Li, X.-D., van den Heuvel, E. P. J., \& Wang, Z.-R. 1998, \apj, 497, 865

\bibitem[{{Li \& Wang}(1996)}]{lw96}
Li, X.-D. \& Wang, Z.-R. 1996, \aap, 307, L5

\bibitem[{{Li \& Wang}(1998)}]{lw98}
Li, X.-D. \& Wang, Z.-R. 1998, \apj, 500, 935

\bibitem[{{Lin} et~al.(2011)}]{l11}
Lin, J., Rappaport, S., Podsiadlowski, P., Nelson, L., Paxton, B.,
\& Todorov, P. 2011, \apj, 732, 70

\bibitem[{{Liu \& Chen} (2011)}]{lc11}
Liu, W.-M. \& Chen, W.-C, 2011, \mnras, 416, 2285

\bibitem[{{L\"ohmer} et~al.(2005)}]{l05}
L\"ohmer, O., Lewandowski, W., Wolszczan, A., \& Wielebinski, R.
2005, \apj, 621, 388

\bibitem[{{Lorimer} et~al.(1995)}]{l95}
Lorimer, D. R., Lyne, A. G., Festin, L., \& Nicastro, L. 1995, \nat, 376, 393

\bibitem[{{Ma \& Li} (2009)}]{ml09}
Ma, B. \& Li, X.-D. 2009, \apj, 691, 1611

\bibitem[{{Manchester} et~al.(2005)}]{m05}
Manchester, R. N., Hobbs, G. B., Teoh, A., \& Hobbs, M. 2005, \aj, 129, 1993

\bibitem[{{Matt \& Pudritz}(2005)}]{mp05}
Matt, S. \& Pudritz, R. E. 2005, \apj, 632, L135

\bibitem[{{Maxted} et~al.(2006)}]{m06}
Maxted, R. F. L., Napiwotzki, R.,  Dobbie, P. D.,  \&  Burleigh, M. R.
2006, \nat, 442, 543 

\bibitem[{{Menou} et~al.(1999)}]{m99}
Menou, K., Esin, A. A., Narayan, R., Garcia, M. R., Lasota, J.-P., \&
McClintock, J. E. 1999, \apj, 520, 276

\bibitem[{{Miller \& Hamilton}(2001)}]{mh01}
Miller, M. C. \& Hamilton, D. P. 2001, \apj, 550, 863

\bibitem[{{Nelemans \& Tauris}(1998)}]{nt98}
Nelemans, G. \& Tauris, T. M. 1998, \aap, 335, L85

\bibitem[{{Nomoto}(1984)}]{n84}
Nomoto, K. 1984, \apj, 277, 791

\bibitem[{{\"Ozel} et~al.(2012)}]{o12}
\"Ozel, F., Psaltis, D., Narayan, R., \& Villarreal, A. S. 2012, ApJ submitted (arXiv:1201.1006)

\bibitem[{{Paczynski} et~al.(1976)}]{p76}
Paczynski, B. 1976, In Structure and Evolution in Close Binary Systems. Proc. IAU
Symp. 73, Eds., Eggleton, P. P., Mitton, S., Whealan, J. (Reidel, Dordrecht), p. 75

\bibitem[{{Papaloizou \& Pringle}(1978)}]{pp78}
Papaloizou, J. \& Pringle, J. E. 1978, \mnras, 184, 501

\bibitem[{{Patruno} et~al.(2012)}]{phd12}
Patruno, A., Haskell, B., \& D'Angelo, C. 2012, \apj, 746, 9

\bibitem[{{Pfahl} et~al.(2003)}]{prp03}
Pfahl, E., Rappaport, S., \& Podsiadlowski, P. 2003, \apj, 597, 1036

\bibitem[{{Podsiadlowski \& Rappaport}(2000)}]{p00}
Podsiadlowski, Ph. \& Rappaport, S. 2000, \apj, 529, 946

\bibitem[{{Podsiadlowski } et~al.(2002)}]{prp02}
Podsiadlowski, Ph., Rappaport, S., \& Pfahl, E. D. 2002, \apj, 565, 1107

\bibitem[{{Pols } et~al.(1995)}]{p95}
Pols, O.R., Tout, C.A., Eggleton, P.P., \& Han Z., 1995, \mnras, 274, 964

\bibitem[{{Pylyser \& Savonije}(1988)}]{ps88}
Pylyser, E. \& Savonije, G. J. 1988, \aap, 191, 57

\bibitem[{{Pylyser \& Savonije}(1989)}]{ps89}
Pylyser, E. \& Savonije, G. J. 1989, \aap, 208, 52

\bibitem[{{Quaintrell} et~al.(2003)}]{q03}
Quaintrell, H., Norton, A. J., Ash, T. D. C., Roche, P., \& Willems, B.
et al. 2003, \aap, 401, 313

\bibitem[{{Rappaport} et~al.(1995)}]{r95}
Rappaport, S.A., Podsiadlowski, P., Joss, P.C., Di Stefano, R., \& Han, Z.
1995, \mnras, 273, 731

\bibitem[{{Rawls} et~al.(2011)}]{r11}
Rawls, M. L., Orosz, J. A., McClintock, J. E., Torres, M. A. P.,
\& Bailyn, C. D. et al. 2011, \apj, 730, 25

\bibitem[{{Refsdal \& Weigert}(1971)}]{rw71}
Refsdal, S. \& Weigert, A. 1971, \aap, 13, 367

\bibitem[{{Ricker \& Taam}(2012)}]{rt12}
Ricker, P. M. \& Taam, R. E. 2012, \apj, 746, 74

\bibitem[{{Ritter}(2008)}]{r08}
Ritter, A. 2008, \nar, 51, 869

\bibitem[{{Romanova} et~al.(2009)}]{r09}
Romanova, M. M., Ustyugova, G. V., Koldoba, A. V., \& Lovelace, R. V. E. 2009,
\mnras, 399, 1802

\bibitem[{{Ruderman} et~al.(1989)}]{rst89}
Ruderman, M., Shaham, J., \& Tavani, M. 1989, \apj, 336, 507

\bibitem[{{Schwab, Podsiadlowski,  \& Rappaport}(2010)}]{spr10}
Schwab, J., Podsiadlowski, P., \& Rappaport, S. 2010, \apj, 719,
722

\bibitem[{{Shao \& Li}(2012)}]{sl12}
Shao, Y. \& Li, X.-D. 2012, \apj, 745, 165

\bibitem[{{Soberman} et~al.(1997)}]{spv97}
Soberman, G. E., Phinney, E. S., \& van den Heuvel, E. P. J. 1997, \aap, 327, 620

\bibitem[{{Stairs} et~al.(2005)}]{s05}
Stairs, I. H., Faulkner, A. J., Lyne, A. G., Kramer, M., \& Lorimer, D. R. et al. 2005, \apj, 632,1060

\bibitem[{{Sutantyo \& Li}(2000)}]{sl00}
Sutantyo, W. \& Li, X.-D. 2000, \aap, 360, 633

\bibitem[{{Tauris}(1996)}]{t96}
Tauris, T. M. 1996. \aap, 315, 453

\bibitem[{{Tauris}(2012)}]{t12}
Tauris, T. M. 2012, Science, 335, 561

\bibitem[{{Tauris}\ et~al.(2011)}]{tlk11}
Tauris, T. M., Langer N., \& Kramer M. 2011, \mnras, 416, 2130

\bibitem[{{Tauris \& Savonije}(1999)}]{ts99}
Tauris, T. M., \& Savonije, G. J. 1999, \aap, 350, 928

\bibitem[{{Tauris \& van den Heuvel}(2006)}]{tv06}
Tauris, T. M. \& van den Heuvel E. P. J.,  2006, in Compact Stellar 
X-Ray Sources, Eds., W. H G. Lewin, M. van der Klis
(Cambridge: Cambridge University Press), 623

\bibitem[{{Tauris}\ et~al.(2000)}]{tvs00}
Tauris, T. M., van den Heuvel, E. P. J., \& Savonije, G. J. 
2000, \apj, 530, L93

\bibitem[{{The Fermi LAT Collaboration}(2011)}]{f11}
The Fermi LAT Collaboration, 2011, Science, 334, 1107

\bibitem[{{Thorsett \& Chakrabarty}(1999)}]{tc99}
Thorsett, S. E. \& Chakrabarty, D. 1999, \apj, 512, 288

\bibitem[{{Timmes} et~al.(1996)}]{tww96}
Timmes, F. X., Woosley, S. E., \& Weaver, Thomas A. 1996, \apj, 457, 834

\bibitem[{{Ustyugova} et~al.(2006)}]{u06}
Ustyugova, G. V., Koleoba, A. V., Romanova, M. M., \& Lovelace, R. V. E. 2006,
\apj, 646, 304

\bibitem[{{van den Heuvel \& Taam}(1984)}]{vt84}
van den Heuvel, E. P. J. \& Taam, R. 1984, \nat, 309, 235

\bibitem[{{van der Meer}et~al.(2007)}]{v07}
van der Meer, A., Kaper, L., van Kerkwijk, M. H., Heemskerk, M. H. M.,
\& van den Heuvel, E. P. J. 2007, \aap, 473, 523

\bibitem[{{van der Sluys, Verbunt, \& Pols}(2005)}]{vs05}
van der Sluys, M. V., Verbunt, F., \& Pols, O. R. 2005, \aap, 440, 973

\bibitem[{{van Paradijs}(1996)}]{v96}
van Paradijs, J. 1996, \apj, 464, L139

%\bibitem[{{Vanbeveren}et~al.(1998)}]{v98}
%Vanbeveren, D., Van Rensbergen, W., \& De Loore, C., 1998b, in The Brightest Binaries,
%Kluwer Academic Pub., Dordrecht

\bibitem[{{Verbunt \& Zwaan}(1981)}]{vz81}
Verbunt, F. \& Zwaan, C. 1981, \aap, 100, L7

\bibitem[{{Wang}(1995)}]{w95}
Wang, Y.-M. 1995, \apj, 449, L153

\bibitem[{{Webbink}(1984)}]{w84}
Webbink, R. F. 1984, \apj, 277, 355

\bibitem[{{Webbink} et~al.(1983)}]{w83}
Webbink, R. F., Rappaport, S. A., \& Savonije, G. J. 1983, \apj, 270, 678

\bibitem[{{Woosley \& Janka}(2005)}]{wj05}
Woosley, S. \& Janka, T. 2005, Nat. Phys., 1, 147

\bibitem[{{Zhang} et~al.(2011)}]{z11}
Zhang, C. M., Wang, J., Zhao, Y. H., Yin, H. X., \& Song, L. M. 
et al. 2011, \aap, 527, A83

\bibitem[{{Zhang \& Li}(2010)}]{zl10}
Zhang, Z. \& Li, X.-D. 2010, \aap, 518, A19

%\bibitem[{{Zorotovic} et~al.(2010)}]{z10}
%Zorotovic, M., Schreiber, M. R., G\"ansicke, B. T., \& Nebot
%G\'omez-Mor\'an, A. 2010, \aap, 520, A86

\end{thebibliography}
\end{document}